# Bayesian inference of mixed Gaussian phylogenetic models

Bayu Brahmantio[1*], Krzysztof Bartoszek[1] and Etka Yapar[2]

*Correspondence:
Bayu Brahmantio
bayu.brahmantio@liu.se
[1]Department of Computer and Information Science, Linköping University, Linköping, Sweden
[2]Department of Biology, Lund University, Lund, Sweden

## Abstract

**Background** Continuous traits evolution of a group of taxa that are correlated through a phylogenetic tree is commonly modelled using parametric stochastic differential equations to represent deterministic change of traits through time, while incorporating noises that represent different unobservable evolutionary pressures. A heterogeneous Gaussian process that consists of multiple parametric sub-processes is often used when the observed data come from a very diverse set of taxa. In the maximum likelihood setting, challenges arise when exploring the involved likelihood surface and when interpreting the uncertainty around the parameters.

**Results** We extend the methods to tackle inference problems for mixed Gaussian phylogenetic models (MGPMs) by implementing a Bayesian scheme that can take into account biologically relevant priors. The posterior inference method is based on the Population Monte Carlo (PMC) algorithm that is easily parallelized, and uses an efficient algorithm to calculate the likelihood of phylogenetically correlated observations. A model evaluation method that is based on the proximity of the posterior predictive distribution to the observed data is also implemented. Simulation study is done to test the inference and evaluation capability of the method. Finally, we test our method on a real-world dataset.

**Conclusion** We implement the method in the R package bgphy, available at https://github.com/bayubeta/bgphy. Simulation study demonstrates that the method is capable to infer parameters to evaluate different models, while its implementation on the real-world dataset indicates that a carefully selected model of evolution based on naturally occurring classifications results in a better fit to the observed data.

## Background

The study of between-species phenotypic data requires taking into consideration the evolutionary dependency structure between them. Approaches for this have been developed from the second-half of the previous century (see, e.g., [1] and [2]). The current widespread availability of computational power and genetically derived phylogenies allowed for an extraordinary growth in numbers of models and software for this field termed phylogenetic comparative methods (PCMs). The high-level framework is that of a branching Markov process, with different processes and parameters corresponding to different biological hypotheses about the evolution of considered traits. However, the majority of contemporary software assumes that the model is homogeneous over the phylogeny, with perhaps some particular parameter being allowed to vary. This is despite evolutionary relationships spanning through millions of years through varying environments. To (at least partially) remedy this, the mixed Gaussian phylogenetic models (MGPMs) framework was introduced [3].

In the MGPM approach, the tree is partitioned into multiple disjoint components (called regimes). Inside each component, the trait(s) evolve under some model from the so-called $\mathcal{G}_{LInv}$ family [4]. An illustration on how the evolutionary processes happen using this framework can be seen on Fig. 1. Under this family the transition density of the trait along a time interval is Gaussian with expectation depending linearly on the ancestral value and a variance that is invariant with respect to the ancestral value. The phylogenetic Brownian motion (BM) and Ornstein-Uhlenbeck (OU) processes, which are the current work-horses of PCMs, are part of this family. Importantly for the inference, this MGPM framework allows for linear in time with respect to the number of tips on the tree likelihood evaluation [4].

The majority of parameter inference techniques for PCMs are maximum likelihood ones, with competing hypotheses concerning the evolutionary history (models, constraints on parameters, regimes on phylogeny configuration) compared using some information criterion. Another view of statistical modelling is to utilize Bayesian methods to infer the posterior distribution of evolutionary parameters given the data.

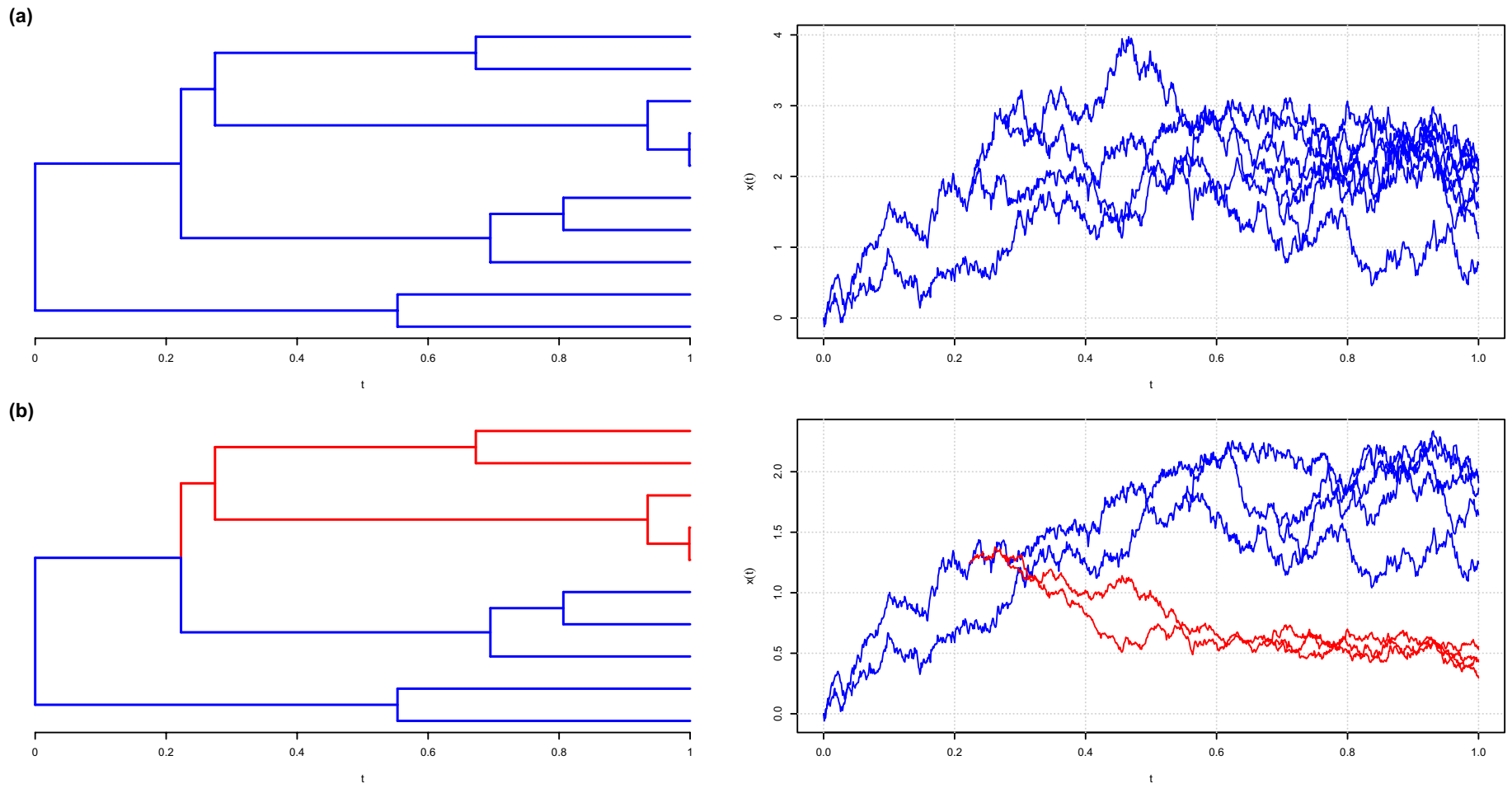

**Fig. 1** An illustration of MGPMs on a simple phylogenetic tree. The right column shows simulated processes that follow the phylogenetic tree on the left. In (**a**), a homogeneous process is assumed throughout the whole tree. In (**b**), the tree is partitioned into the blue and red regimes, each with its own process. The evolution starts with the process in the blue regime at time $t = 0$, but one part of the tree switches to the process in the red regime at some point in time until $t = 1$

In this work, we enhance the framework of MGPMs by implementing a Bayesian inference method for the combinations of univariate BM and OU processes on the phylogeny. We apply a Monte Carlo method that is based on the importance sampling technique for simulating from the posterior distribution of the model's parameters. Furthermore, we implement a model evaluation method that takes into account the proximity between the empirical distribution and the posterior predictive distribution.

The rest of the paper is organized as follows. We start by explaining the notation and theoretical background needed for the rest of the paper. In the following subsection, we discuss the Bayesian approach to MGPMs. We proceed by describing our method to simulate the posterior distribution of MGPM parameters and the software that we implemented. We then explain the setup for the simulation studies and implement our method into antlers dataset [5]. The results of the simulation studies and the real-world dataset are discussed, as well as its limitations. We conclude our paper by discussing the results and future improvements of our work.

## Notations and theoretical background

### *Phylogenetic likelihood of continuous traits*

We denote $\mathrm{x} = (x_1, x_2, \cdots, x_n)^\intercal \in \mathbb{R}^n$ as the vector of continuous traits measurements where each $x_i$ denotes a trait for taxon *i*. As examples, $\mathrm{x}$ could be the body size of different species of lizards [6], the length cichlid species [7], or the freezing index of injury at -15°C of oak species [8]. Since one species can have multiple data points, $x_i$ is usually set to be the mean trait of the species.

A "naïve" approach to fit a statistical distribution to the observed data $\mathrm{x}$ is to consider $x_1, x_2, \cdots, x_n$ as independent samples of the same distribution *X*, which has a probability density function (PDF) $p_X(x|\boldsymbol{\Theta})$, with $\boldsymbol{\Theta}$ as the parameter. This leads to a straightforward calculation of the likelihood function, which is given by

$$p(\mathrm{x}|\boldsymbol{\Theta}) = \prod_{i=1}^{n} p_X(x_i|\boldsymbol{\Theta}). \tag{1}$$

However, the assumption of independence ignores the fact that the taxa are related to one another because they have a shared ancestry, explained by a phylogenetic tree.

We denote $\mathcal{T}$ as an ultrametric phylogenetic tree that governs the relationship between the traits $x_1, x_2, \cdots, x_n$. This means that $\mathcal{T}$ has *n* number of tips, where tip *i* corresponds to the taxon *i*. Consequently, $\mathcal{T}$ has $n-1$ internal nodes and $2n-2$ edges (or branches). To further elucidate this, we can take an example from the phylogenetic tree in Fig. 1, where it has 10 tips, 9 internal nodes, and 18 branches.

We assume that the tree is known and fixed and that the data can only be observed at the tips. That is, the trait $x_i$ is observed at the tip *i*. Furthermore, we denote $\mathrm{z} = (z_1, z_2, \cdots, z_{n-1})^\intercal$ as the internal nodes of $\mathcal{T}$. The set of edges of the tree $\mathcal{T}$ can be written as $\mathrm{e}(\mathcal{T})$. Together, $\mathrm{x}$, $\mathrm{z}$, and $\mathrm{e}(\mathcal{T})$ completely determine the tree $\mathcal{T}$.

Assuming that *X* follows a branching Markov process, the likelihood function in Equation (1) under the phylogenetic relationship $\mathcal{T}$ becomes

$$p(\mathrm{x}|\boldsymbol{\Theta}, \mathcal{T}) = \prod_{v \in V} p_X(v|\mathrm{pa}(v), \boldsymbol{\Theta}, \mathcal{T}), \tag{2}$$

where $V = \{x_1, x_2, \cdots, x_n, z_1, z_2, \cdots, z_{n-1}\}$, i.e., the combined set of tips and internal nodes, and $\mathrm{pa}(v)$ denotes the parent node of node *v*. We can also see PDF $p_X(v|\mathrm{pa}(v), \boldsymbol{\Theta}, \mathcal{T})$ as a transition PDF of *v* given an ancestral value $\mathrm{pa}(v)$.

From a macroevolutionary perspective, we can model changes in traits over time using stochastic differential equations (SDE). Two of the most popular models of SDEs are Brownian motion (BM) and Ornstein-Uhlenbeck (OU) processes, which can be written respectively as

$$dx(t) = \sigma dW(t), \tag{3}$$

and

$$dx(t) = -\alpha(x(t) - \theta)dt + \sigma dW(t). \tag{4}$$

The BM process can be interpreted as an evolutionary process that spreads from the ancestral value *x*(*s*) without a long-term direction, while the OU process is a stabilizing evolutionary process that converges to an optimum trait. Notice that the OU process has additional parameters $\alpha$ and $\theta$, where $\theta$ is the optimum trait and $\alpha$ determines how fast the trait evolves from *x*(*s*) to $\theta$. If $\alpha = 0$, then the OU process is equivalent to the BM process, while a larger value of $\alpha$ speeds up the convergence of the trait to $\theta$.

Given a trait value at time *s*, the BM process induces the transition PDF

$$x(t)|x(s) \sim \mathcal{N}\left(x(t)\middle|x(s), \sigma^2 l\right), \tag{5}$$

while the OU process induces the transition PDF

$$x(t)|x(s) \sim \mathcal{N}\left(x(t)\middle|e^{-\alpha l}x(s) + (1 - e^{-\alpha l})\theta,\ \frac{\sigma^2}{2\alpha}(1 - e^{-2\alpha l})\right), \tag{6}$$

where $l = t - s$, for $t > s$. Figure 1a, right panel, illustrates a realization trajectory of an OU process on top of a phylogenetic tree.

Since both processes have a normal transition PDF, the observed data at the tips $\mathrm{x}$ follows an *n*-variate normal distribution, given a value at the root $X_0$ (see, e.g., [9]). This implies that calculating the likelihood in Equation (2) requires inversion of the $n \times n$ variance-covariance matrix of $\mathrm{x}$, which has a computational complexity of $\mathcal{O}(n^3)$. To overcome this, the "pruning" algorithm is implemented, which breaks down the likelihood calculation into a dynamic programming problem by recursively calculating the likelihood in a post-order direction of the phylogenetic tree (see [4, 10–12]) with a computational complexity of $\mathcal{O}(n)$.

A general family of models, called $\mathcal{G}_{LInv}$ [4], includes models that have a normal transition PDF, where the mean is linear to the ancestral value, and the variance does not depend on the ancestral value. Specifically, for a model $\mathcal{M}$ with parameters $\Theta$ belonging to the $\mathcal{G}_{LInv}$ family, the conditional mean at time *t* given an ancestral value at time *s*, *x*(*s*), can be written as

$$\mathbb{E}[x(t)|x(s)] = a_{l,\mathcal{M},\Theta}x(s) + b_{l,\mathcal{M},\Theta}, \tag{7}$$

for some constants $a_{l,\mathcal{M},\Theta}$ and $b_{l,\mathcal{M},\Theta}$, and the variance is

$$\mathrm{Var}(x(t)|x(s)) = V_{l,\mathcal{M},\Theta}. \tag{8}$$

This suggests that the BM and OU processes are in the $\mathcal{G}_{LInv}$ family, where it can be shown that, for the BM case,

$$a_{l,BM,\sigma} = 1, \tag{9}$$

$$b_{l,BM,\sigma} = 0, \tag{10}$$

$$V_{l,BM,\sigma} = \sigma^2 l, \tag{11}$$

while for the OU case,

$$a_{l,OU,\{\alpha,\theta,\sigma\}} = e^{-\alpha l}, \tag{12}$$

$$b_{l,OU,\{\alpha,\theta,\sigma\}} = (1 - e^{-\alpha l})\theta, \tag{13}$$

$$V_{l,OU,\{\alpha,\theta,\sigma\}} = \frac{\sigma^2}{2\alpha}(1 - e^{-2\alpha l}). \tag{14}$$

The likelihood calculations with the pruning algorithm for the $\mathcal{G}_{LInv}$ family are handled by the `PCMBase` with an additional speed-up using `C++` backend with `PCMBaseCPP`.

***Evolutionary regimes***

Until now, we assumed that $\mathrm{x}$ was the result of a single type of process (BM or OU). We now introduce the "evolutionary regimes" (or "regimes" shortly) as a way to utilize several models of evolution on top of the phylogenetic tree. We assume a coloring set of the edges $\mathrm{e}(\mathcal{T})$ of $\mathcal{T}$, which we will address as "regimes configuration", that results in $K$ disjoint sets of $\mathrm{e}(\mathcal{T})$. For example, in Fig. 1b, each edge in $\mathrm{e}(\mathcal{T})$ belongs to the "red" or "blue" regime, but not both. This setup also makes it possible for a regime to be disconnected across different parts of the tree.

Using the $\mathcal{G}_{LInv}$ framework, denote $\mathrm{M} = \{\mathcal{M}_i\}_{i=1}^K$ as the set of all the models from all the $K$ regimes, with the corresponding parameters $\mathbf{\Theta} = \{\Theta_i\}_{i=1}^K$. Since all models are in the $\mathcal{G}_{LInv}$ family, the transition probability under each model, given some ancestral value $x(s)$, $s < t$, is

$$\left(x(t) \,\middle|\, x(s), \mathcal{M}_i, \Theta_i\right) \sim \mathcal{N}\left(x(t) \,\middle|\, a_{l,\mathcal{M}_i,\Theta_i} x(s) + b_{l,\mathcal{M}_i,\Theta_i},\ V_{l,\mathcal{M}_i,\Theta_i}\right), \tag{15}$$

where $l = t - s$, $\mathcal{M}_i$ is the model type (BM or OU) for the regime $i$ with parameters $\Theta_i$. We can also set the ancestral value at the root of the tree, $X_0$, as a free parameter. Hence, the full collection of parameters from a given MGPM with $K$ regimes is $\mathbf{\Theta} = (X_0, \Theta_1, \cdots, \Theta_K)^\intercal = (X_0, \vartheta_{1,1}, \cdots, \vartheta_{1,n_1}, \cdots, \vartheta_{K,1} \cdots \vartheta_{K,n_K})^\intercal$.

As an example, consider an MGPM with $K = 2$, $\mathcal{M}_1(\Theta_1) = \mathrm{BM}(\sigma_1)$, and $\mathcal{M}_2(\Theta_2) = \mathrm{OU}(\alpha_2, \theta_2\sigma_2)$. Then, the full collection of models for this MGPM is $\mathrm{M} = \{\mathrm{BM}, \mathrm{OU}\}$ with the full parameters $\boldsymbol{\Theta} = (X_0, \sigma_1, \alpha_2, \theta_2, \sigma_2)^\intercal$.

**Bayesian approach to MGPMs**

In the maximum likelihood setting, the goal is to find the parameters $\boldsymbol{\Theta}^*$ that maximize the likelihood function $\mathcal{L}(\boldsymbol{\Theta})$, for a given MGPM (e.g., [3]). Additionally, we can calculate their information score, e.g., AIC [13] or BIC [14], to choose between different hypothesized models.

Another perspective to tackle the parameter estimation and model evaluation problem is to use a Bayesian approach. As discussed in previous studies [15–17], a relevant prior distribution of the parameters helps to alleviate the identifiability issues that exist in the OU models, which is one of the models considered in this study. A prior distribution is also a natural way to safeguard inference from unrealistic parameter values that could arise in the maximum likelihood setting, where the maximum search could be stuck in a local optimum that does not always translate to a realistic parameter value. Furthermore, the posterior distribution offers an inherent probabilistic interpretation of the uncertainty of the parameters.

Applying Bayes' theorem, the density function of the posterior distribution of the parameters of a given MGPM setup is

$$p\left(\boldsymbol{\Theta}|\mathrm{x}\right) = \frac{p\left(\mathrm{x}|\boldsymbol{\Theta}\right)p\left(\boldsymbol{\Theta}\right)}{p\left(\mathrm{x}\right)} \propto \mathcal{L}(\boldsymbol{\Theta})p(\boldsymbol{\Theta}), \tag{16}$$

where the parameters are assumed to be pairwise independent in the prior, i.e.,

$$p(\boldsymbol{\Theta}) = p(X_0)\prod_{i=1}^{K} p(\Theta_i) = p(X_0)\prod_{i=1}^{K}\prod_{j=1}^{n_K} p(\vartheta_{i,j}), \tag{17}$$

and we drop the notation of the implicit dependence on the model M, the tree $\mathcal{T}$, and the distribution $X$. Hence, we can individually set the appropriate prior distribution for each parameter. The posterior distribution is in general not available in a closed-form formula. Since we can still compute the likelihood and the prior density, we resort to an approximation procedure with Monte Carlo methods to calculate quantities of the posterior distribution.

This work differs from previously done studies on Bayesian PCMs in different ways. In `bayou` [15], the regimes are considered random, and reversible-jump MCMC is used to simulate the posterior distribution. However, some parameters of the OU processes defined for the regimes, namely $\alpha$ and $\sigma$, are common for all regimes, while $\theta$ can vary across the regimes. The ancestral value at the root $X_0$ is also set to be the same as one of the $\theta$ values. Another method [18] uses Hamiltonian Monte Carlo to simulate the posterior distribution of the OU model for multivariate observations at the tips, but does not allow for different regimes on the tree. Although we assume that the regimes are fixed and the value at the tips to be univariate, we allow more flexibility in defining the models by allowing each regime to have either BM or OU process with complete parameters. The only common parameter for all regimes is the ancestral value at the root $X_0$. We also implemented a Monte Carlo algorithm that is based on importance sampling, which could be easily parallelized since it involves a lot of independent calculations.

## Methods

We begin this section by explaining the procedures behind the method that we implemented, starting from the posterior sampling algorithm and the mechanism to evaluate the model. We then briefly explain the software implementation of our method. We proceed by explaining the setup for simulation study to assess the performance of our method in terms of parameter inference and model evaluation. Lastly, we describe the real-world data that we use and postulate different scenarios of evolution, to be tested using our method.

### Bayesian inference of MGPM parameters

For convenience, the observation vector at the tips, $\mathrm{x} = (x_1, x_2, \cdots, x_n)^\intercal$, is assumed to take values from $\mathbb{R}^n$. Since the observations are generally positive-valued, a one-to-one transformation to $\mathbb{R}^n$ using the log function is commonly used before the inference procedure.

We implement a procedure that consists of multiple steps to simulate the posterior distribution of $\boldsymbol{\Theta}$ given an MGPM $\mathbb{M}$ with $K$ regimes. First, we define the prior distribution of $\boldsymbol{\Theta}$ by defining a prior distribution for each parameter individually, as explained in Equation (17). The choices of different prior distributions are available in the `bgphy` package.

To make things easier for the Monte Carlo algorithm that involves sampling from the multivariate normal distribution, we transform the prior distribution into a transformed prior distribution that takes values from $\mathbb{R}^d$, where $d = \dim(\boldsymbol{\Theta})$. Since the parameters are assumed to be pairwise independent, we use a class of well-behaved and one-to-one functions [19] to transform the parameters individually. Let $g : \mathcal{S}_{\boldsymbol{\Theta}} \to \mathbb{R}^d$, where $\mathcal{S}_{\boldsymbol{\Theta}}$ is the original space of the parameters $\boldsymbol{\Theta}$.

With a slight abuse of notation, we keep the symbol $\boldsymbol{\Theta}$ for the transformed prior, but we remind the reader that we are working on the transformed, unbounded space of $\mathbb{R}^d$ that is the result of transforming the original $\boldsymbol{\Theta}$ space by the function $g$, i.e., $\boldsymbol{\Theta} \leftarrow g(\boldsymbol{\Theta})$.

The posterior sampling scheme follows the Population Monte Carlo (PMC, [20]) procedure, which involves several schemes of importance sampling and resampling. The PMC scheme is chosen because it is an improvement over the standard importance sampling scheme, which is heavily dependent on the choice of the proposal distribution. On the other hand, it can be seen as a specific case of a more advanced Sequential Monte Carlo sampler scheme [21], but requires fewer likelihood calculations which can be relatively expensive because of the phylogenetic structure behind the observations. In this manner, PMC is a good trade-off between inference quality and speed.

We adapt the PMC algorithm by using a normal distribution obtained from Laplace's approximation (see, e.g., [22]) as the initial generating distribution and local normal distributions as the generating distributions after a resampling step.

In the Laplace's approximation step, we approximate the posterior distribution with a multivariate normal distribution. Specifically, we match the mean of the normal distribution with the posterior mode, $\hat{\boldsymbol{\Theta}}$, and match the covariance of the normal distribution with $(-\hat{\mathrm{H}}_{\hat{\boldsymbol{\Theta}}})^{-1}$, where $\hat{\mathrm{H}}_{\hat{\boldsymbol{\Theta}}}$ is the approximate Hessian matrix around the posterior mode, $\hat{\boldsymbol{\Theta}}$. We denote the density function of the resulting multivariate normal distribution as $q(\boldsymbol{\Theta})$. Then, we draw samples from this multivariate normal distribution,

$$\boldsymbol{\Theta}_1^{(1)}, \cdots, \boldsymbol{\Theta}_S^{(1)} \overset{iid}{\sim} \mathcal{MVN}\left(\boldsymbol{\Theta} \middle| \hat{\boldsymbol{\Theta}}, (-\hat{\mathrm{H}}_{\hat{\boldsymbol{\Theta}}})^{-1}\right), \tag{18}$$

We then calculate the (normalized) weights of the drawn samples by

$$w_i^{(1)} = \frac{\tilde{w}(\boldsymbol{\Theta}_i^{(1)})}{\sum_{j=1}^{S} \tilde{w}(\boldsymbol{\Theta}_j^{(1)})}, \text{ where } \tilde{w}(\boldsymbol{\Theta}_i^{(1)}) = \frac{\mathcal{L}(\boldsymbol{\Theta}_i^{(1)}) p(\boldsymbol{\Theta}_i^{(1)})}{q(\boldsymbol{\Theta}_i^{(1)})}, \tag{19}$$

where $\mathcal{L}(\cdot)$ is the likelihood function and $p(\cdot)$ is the prior PDF. In the next step, a resampling step is performed by creating a new set of samples from the old ones. We do this by randomly choosing $\boldsymbol{\Theta}_1^{(1)}, \cdots, \boldsymbol{\Theta}_S^{(1)}$ with replacements and with probabilities $w_1^{(1)}, \cdots, w_S^{(1)}$. In other words, we sample indices

$$a_1, \cdots, a_S \overset{iid}{\sim} \text{Categorical}(\{w_1^{(1)}, \cdots, w_S^{(1)}\}), \tag{20}$$

where $a_1, \cdots, a_S$ are i.i.d categorical with probabilities $w_1^{(1)}, \cdots, w_S^{(1)}$. This is also known as the multinomial resampling scheme in other literature (e.g., [23]). This step is important in throwing away samples that have low or close-to-zero weights, which indicates that they are far from the high posterior density area.

We proceed by drawing new samples $\boldsymbol{\Theta}_1, \cdots, \boldsymbol{\Theta}_S$, where

$$\boldsymbol{\Theta}_i \sim \mathcal{MVN}\left(\boldsymbol{\Theta} \middle| \boldsymbol{\Theta}_i^{(2)}, \boldsymbol{\Sigma}_{\boldsymbol{\Theta}^{(2)}}\right). \tag{21}$$

That is, for each sample $\boldsymbol{\Theta}_i^{(2)}$, we draw a new sample from a multivariate normal distribution centered around $\boldsymbol{\Theta}_i^{(2)}$, while the covariance matrix can be made common for all $i = 1, \cdots, S$. We set $\boldsymbol{\Sigma}_{\boldsymbol{\Theta}^{(2)}}$ as a diagonal matrix where $\boldsymbol{\Sigma}_{\boldsymbol{\Theta}^{(2)}} = \frac{1}{S}\text{diag}(\hat{\sigma}_1^2, \cdots, \hat{\sigma}_d^2)$, i.e., $\frac{1}{S}$ times a diagonal matrix where the element in the *j*-th row and column is the empirical variance of the *j*-th dimension of $\boldsymbol{\Theta}^{(2)}$ from all *S* samples, $1 \leq j \leq d$.

Finally, we calculate the weights for the new sample with a procedure similar to that of Equation (19), but we change the density function $q(\cdot)$ into local densities $q_i(\cdot)$, so that

$$w_i = \frac{w^*(\boldsymbol{\Theta}_i)}{\sum_{j=1}^{S} w^*(\boldsymbol{\Theta}_j)}, \text{ where, } w^*(\boldsymbol{\Theta}_i) = \frac{\mathcal{L}(\boldsymbol{\Theta}_i) p(\boldsymbol{\Theta}_i)}{q_i(\boldsymbol{\Theta}_i)}, \tag{22}$$

where $q_i$ is the density function of the multivariate normal distribution that generates $\boldsymbol{\Theta}_i$. The pseudo-code for the posterior sampling scheme is illustrated by Algorithm 1.

**Require**:

- Phylogenetic tree $\mathcal{T}$.
- Observed data at the tips, $\mathbf{x}$.
- An MGPM $\mathbf{M}$, with parameters set $\mathbf{\Theta}$, $\dim(\mathbf{\Theta}) = d$, consisting of a regime configurations with $K$ regimes.
- Prior distribution $p(\mathbf{\Theta})$.
- Number of Monte Carlo samples, $S$.
- Transformation function, $g$.

**Output**: Pairs of samples from the parameter space and their weights, $\{\mathbf{\Theta}_1, w_1\}, \cdots, \{\mathbf{\Theta}_S, w_S\}$.

**Begin**:

Transform $\mathbf{\Theta}$ and $p(\mathbf{\Theta})$ $g$.

$\mathbf{\Theta}_1^{(1)}, \cdots, \mathbf{\Theta}_S^{(1)} \overset{iid}{\sim} q(\mathbf{\Theta}) = \mathcal{MVN}\left(\mathbf{\Theta} \middle| \hat{\mathbf{\Theta}}, (-\hat{\mathbf{H}}_{\hat{\mathbf{\Theta}}})^{-1}\right)$

**for** $i = 1, \cdots, S$ **do**

  $w_i^{(1)} = \frac{\tilde{w}(\mathbf{\Theta}_i^{(1)})}{\sum_{j=1}^{S} \tilde{w}(\mathbf{\Theta}_j^{(1)})}$, where $\tilde{w}(\mathbf{\Theta}_i^{(1)}) = \frac{\mathcal{L}(\mathbf{\Theta}_i^{(1)})p(\mathbf{\Theta}_i^{(1)})}{q(\mathbf{\Theta}_i^{(1)})}$.

**end for**

**for** $i = 1, \cdots, S$ **do**

  $a_i \sim \text{Categorical}(\{w_1^{(1)}, \cdots, w_S^{(1)}\})$

  $\mathbf{\Theta}_i^{(2)} = \mathbf{\Theta}_{a_i}^{(1)}$

**end for**

$\mathbf{\Sigma}_{\mathbf{\Theta}^{(2)}} = \frac{1}{S}\text{diag}(\hat{\sigma}_1^2, \cdots, \hat{\sigma}_d^2)$, where $\hat{\sigma}_j^2$ is the sample variance of the $j$-th dimension of $\mathbf{\Theta}^{(2)}$ from all $S$ samples, $1 \leq j \leq d$.

**for** $i = 1, \cdots, S$ **do**

  $\mathbf{\Theta}_i \sim q_i(\mathbf{\Theta}) = \mathcal{MVN}\left(\mathbf{\Theta} \middle| \mathbf{\Theta}_i^{(2)}, \mathbf{\Sigma}_{\mathbf{\Theta}^{(2)}}\right)$

**end for**

**for** $i = 1, \cdots, S$ **do**

  $w_i = \frac{w^*(\mathbf{\Theta}_i)}{\sum_{j=1}^{S} w^*(\mathbf{\Theta}_j)}$, where $w^*(\mathbf{\Theta}_i) = \frac{\mathcal{L}(\mathbf{\Theta}_i)p(\mathbf{\Theta}_i)}{q_i(\mathbf{\Theta}_i)}$.

**end for**

Transform back $\{\mathbf{\Theta}_i\}_{i=1}^{S}$ using $g^{-1}$.

**Algorithm 1** Posterior sampling scheme

The output of this sampling procedure is a set of pairs of parameter values and their weights $\{\mathbf{\Theta}_1, w_1\}, \cdots, \{\mathbf{\Theta}_S, w_S\}$, which can be directly used to approximate the integrals that arise when calculating the expected values of functions with respect to the posterior distribution:

$$\mathbb{E}_{\mathbf{\Theta}|\mathrm{x}}[h(\mathbf{\Theta})] = \int h(\mathbf{\Theta}) p\left(\mathbf{\Theta}|\mathrm{x}\right) d\mathbf{\Theta} \approx \sum_{i=1}^{S} h(\mathbf{\Theta}_i) w_i. \tag{23}$$

As an example, $h(\mathbf{\Theta}) = \mathbf{\Theta}$ results in the expected value of the posterior distribution, $\mathbb{E}_{\mathbf{\Theta}|\mathrm{x}}[\mathbf{\Theta}]$, which can be approximated by the sum $\sum_{i=1}^{S} \mathbf{\Theta}_i w_i$.

## Model evaluation and comparison

The conventional method to evaluate models in a Bayesian setting is to use the Bayes factor [24], where the ratio of marginal likelihood terms of competing models $\mathrm{M}_0$ and $\mathrm{M}_1$,

$$BF_{1,0} = \frac{p(\mathrm{x}|\mathrm{M}_1)}{p(\mathrm{x}|\mathrm{M}_0)}, \tag{24}$$

is calculated. This is viewed as a measure of the evidence found in the observed data for one statistical model versus another. Each term in the above ratio can be obtained by computing the integral

$$p(\mathrm{x}|\mathrm{M}_i) = \int p(\mathrm{x}|\boldsymbol{\Theta}_i, \mathrm{M}_i) p(\boldsymbol{\Theta}_i|\mathrm{M}_i) d\boldsymbol{\Theta}_i, \quad (25)$$

for $i = 0, 1$, which implicitly penalizes the complexity of the model, since a more constrained model is preferred when it results in a similar fit to observations as the more complex model [25].

However, the Bayes factor is not without its shortcomings. The Bayes factor in Equation (24) alone is not enough to determine which model is more likely to be true given the observed data, which is done by calculating the posterior odds,

$$\frac{p(\mathrm{M}_1|\mathrm{x})}{p(\mathrm{M}_0|\mathrm{x})} = \frac{p(\mathrm{x}|\mathrm{M}_1)}{p(\mathrm{x}|\mathrm{M}_0)} \times \frac{p(\mathrm{M}_1)}{p(\mathrm{M}_0)} \quad (26)$$

$$= BF_{1,0} \times \frac{p(\mathrm{M}_1)}{p(\mathrm{M}_0)}, \quad (27)$$

where the term $p(\mathrm{M}_1)/p(\mathrm{M}_0)$ is the prior odds. By directly calculating Equation (24), we assume that $p(\mathrm{M}_1) = p(\mathrm{M}_0)$. Therefore, the threshold for deciding whether a model can be considered more plausible or not given the observed data depends on the prior beliefs on the models, which are not easy to determine. Moreover, additional computations must be performed to obtain the integral in Equation (25), which is usually not available in closed form. In addition, the marginal likelihood can be sensitive to the choice of the prior over the parameters $\boldsymbol{\Theta}_i$ (see [26] and [27]).

Another way to evaluate the goodness-of-fit of a model to the observed data is by comparing the data replicated under the model and the observed data. In principle, a model that is more plausible for explaining the data-generating process should be able to replicate data at the tips that are close to the observed data. To do this, we compare the posterior predictive distribution with the observed data [28].

In our problem, the density of the posterior predictive distribution can be formulated as

$$p(\tilde{\mathrm{x}}|\mathrm{x}) = \int p(\tilde{\mathrm{x}}|\boldsymbol{\Theta}) p(\boldsymbol{\Theta}|\mathrm{x}) d\boldsymbol{\Theta}, \quad (28)$$

i.e., it is a weighted average of the likelihood function $\mathcal{L}(\boldsymbol{\Theta}) = p(\tilde{\mathrm{x}}|\boldsymbol{\Theta})$ over the posterior distribution $\boldsymbol{\Theta}|\mathrm{x}$.

We adapt the posterior predictive loss formulation [29] to our problem as follows:

$$\begin{aligned} &\|\mathrm{x} - \mathbb{E}[\tilde{\mathrm{x}}|\mathrm{x}]\|_2 + \mathrm{tr}(\mathrm{Cov}(\tilde{\mathrm{x}}|\mathrm{x})) \\ &\quad = \sum_{i=1}^{n} (x_i - \mathbb{E}[\tilde{x}_i|\mathrm{x}])^2 + \sum_{i=1}^{n} \mathrm{Var}(\tilde{x}_i|\mathrm{x}), \end{aligned} \quad (29)$$

where $x_i$ is the observed data point at tip *i* and $\tilde{x}_i|\mathrm{x}$ is the marginal posterior predictive distribution at tip *i*.

The formula above consists of two terms. The first term is the sum of squared errors (SSE) to measure the dissimilarity between the observed data at the tip *i* and the expected value of the marginal posterior predictive distribution at the tip *i*. Thus, a lower SSE score indicates that the posterior predictive distribution is closer to the observed data. The second term is the sum of the variances of the marginal posterior predictive distribution of all tips. Under a model with many parameters, the SSE would be lower, but the sum of variances would be higher since the replicated data could be more diverse. Therefore, the sum of variances term acts as an implicit penalizing term for the complexity of the model.

The posterior predictive loss in Equation (29) is based on the posterior predictive distribution $\tilde{\mathrm{x}}|\mathrm{x}$, which is not analytically available. However, we can still estimate it by simulating samples from the posterior predictive distribution. We do this in two steps: first, we sample parameter values from the posterior distribution and second, we simulate observations at the tips by running the MGPM with the sampled parameter values. Using the results of Sect. 2.1, samples from the posterior distribution can be drawn by performing a multinomial resampling process over the posterior parameters $\mathbf{\Theta}_i, \cdots, \mathbf{\Theta}_S$, with probabilities determined by the final weights $w_1, \cdots, w_S$. The pseudo-code of this posterior predictive sampling process can be found in Algorithm 2.

**Require**:

- Phylogenetic tree $\mathcal{T}$.
- An MGPM $\mathbf{M}$ consisting of a regime configurations with $K$ regimes.
- Number of Monte Carlo samples, $S_p$.
- Parameters and weights from Algorithm 1, $\{\mathbf{\Theta}_1, w_1\}, \cdots, \{\mathbf{\Theta}_S, w_S\}$.

**Output**: Samples from the posterior predictive distribution, $\tilde{\mathbf{x}}_1, \cdots, \tilde{\mathbf{x}}_{S_p}$.
**Begin**:

```
for i = 1, ..., S_p do
    Sample from posterior distribution:
    a_i ~ Categorical({w_1, ..., w_S})
    Θ*_i = Θ_{a_i}
    Simulate observations at the tips:
    x̃^(i) ~ M(Θ*_i, T)
end for
```

**Algorithm 2** Posterior predictive sampling scheme

The simulated observations from the posterior predictive distribution, $\tilde{\mathrm{x}}^{(1)}, \cdots, \tilde{\mathrm{x}}^{(S)}$, from Algorithm 2 can be used to estimate Equation (29). Let $\tilde{\mathrm{x}}^{(j)} = (\tilde{x}_1^{(j)}, \tilde{x}_2^{(j)}, \cdots, \tilde{x}_n^{(j)})^\intercal$ be the *j*-th simulated observation at the tips, where $\tilde{x}_i^{(j)}$ is the simulated observation at tip *i*. Then, the estimated posterior predictive loss is

$$\text{post.}\ \widehat{\text{pred.}}\ \text{loss} = \sum_{i=1}^{n}(x_i - \tilde{m}_i)^2 + \sum_{i=1}^{n}\tilde{s}_i^2, \tag{30}$$

where

$$\tilde{m}_i = \frac{1}{S_p}\sum_{j=1}^{S_p}\tilde{x}_i^{(j)} \quad \text{and} \quad \tilde{s}_i^2 = \frac{1}{S_p - 1}\sum_{j=1}^{S_p}(\tilde{x}_i^{(j)} - \tilde{m}_i)^2. \tag{31}$$

Respectively, $\tilde{m}_i$ and $\tilde{s}_i^2$ are the *i*-th element of the sample mean and the *i*-th element of the diagonal of the sample variance-covariance matrix from all simulated observations.

### Software

Our software is available as an `R` [30] package `bgphy`, which can be downloaded from https://github.com/bayubeta/bgphy. The package contains the main function `bgphy()`, which takes the following arguments: `model`, `X`, `nsample`, and `parallel`. The argument `model` is an object created by the `setModel()` function, which provides information about the phylogeny, regimes, and model types in all regimes. We provide the measurements at the tips through argument `X`. The number of posterior samples, *S*, is provided through `nsample`. Finally, the argument `parallel` is a Boolean that indicates whether parallel computations should be performed or not.

The function `bgphy()` produces a list that contains information on the simulated posterior distribution: mean, standard deviation, standard error for the mean, quantiles (2.5%, 50%, 97.5%), posterior predictive loss, and effective sample size (ESS). The pairs of parameters and their weights, $\{\mathbf{\Theta}_i, w_i\}_{i=1}^{S}$, are also returned so that they can be used to calculate the expected values of arbitrary functions with respect to the posterior distribution (Eq. 23).

In terms of computational complexity, our software runs at $\mathcal{O}(n)$, where *n* is the number of tips. This is due to `PCMBase`'s post-order likelihood calculation scheme. Additionally, we added an option for parallel likelihood computation, as the likelihood calculation can be done independently given different sets of parameters.

In practice, a posterior inference run of an MGPM with two OU regimes (7 parameters) on a tree with 100 tips and 10,000 samples takes around 203.0453 ± 9.7687 s on a personal computer with a 6-core, 12-thread CPU and 16 GB of RAM. Using the same setup, a maximum likelihood fit using `PCMFit` [3], which uses the same `PCMBase` engine, takes around 5.7344 ± 0.8161 s. This is not surprising, as Bayesian inference requires a lot more likelihood calculations. However, it is advisable to perform multiple runs of maximum likelihood fits since the results can vary a lot depending on the starting point. So, a run consisting of 40 maximum likelihood fits takes a similar computational time as one posterior inference using `bgphy` with 10,000 samples.

We also compared the computational time of `bgphy` with `bayou` [15] which uses MCMC. To reduce the autocorrelation between MCMC samples, we retained a sample for every 10 MCMC samples. This means that we needed to run 100,000 MCMC moves to obtain the same number of samples as `bgphy` (10,000). Furthermore, we need to calculate the Bayes factor using a 5-step stepping-stone algorithm. We recorded around 239.6764 ± 17.3111 s for a posterior inference to obtain posterior 10,000 samples.

Ultimately, the comparisons are not entirely impartial, since we compared our method with a maximum likelihood method and another Bayesian method that uses MCMC which can be made quicker given the right configurations. However, we point out that our method is at least on par in terms of computational time with other familiar methods.

### Method assessment

#### *Simulation study*

**Parameter estimation**. To see the capability of our method in terms of parameter estimation, we performed a study based on simulated data. We used the post-processed phylogenetic tree of *Anolis* lizards [6, 31] which has a unit height and simulated data at the tips of the tree. For the MGPM, we created a scenario where there are two regimes on the tree: *Ancestral*, which has an OU process with parameters $\alpha_1 = 2$, $\theta_1 = 2$, and $\sigma_1 = 1$, and $\mathrm{R}_1$ which has an OU process with parameters $\alpha_2 = 5$, $\theta_2 = 0$, and $\sigma_2 = 0.5$. The ancestral value at the root is set to $X_0 = 0$. The regimes configuration on the tree and an example of the process to simulate the observations at the tips can be seen in Fig. 2.

The priors for the parameters are set the same depending on their space. We put half-normal prior with a scale parameter $\sigma = 10$ for the positive parameters, $\alpha_1, \sigma_1, \alpha_2$, and $\sigma_2$. For parameters that can take values from $\mathbb{R}$, $X_0$, $\theta_1$, and $\theta_2$, we chose the normal distribution with $\mu = 0$ and $\sigma^2 = 100$.

Using this setup, we simulated 100 sets of observations at the tips. For each simulated set of observations, we inferred the posterior parameters and compared them with the true parameter values. For comparison, we also performed a maximum likelihood fit given each set of observations using $\mathrm{PCMFit}$ [3]. Because of the random initialization for the maximum likelihood procedure and the complicated likelihood surface, we performed 50 maximum likelihood runs for each set of observations and picked a set of parameters that has the highest log-likelihood. We also restricted the search space to [0, 100] for positive parameters ($\alpha$'s and $\sigma$'s) and $[-200, 200]$ for $X_0$ and $\theta$'s.

For both posterior and maximum likelihood inference, we assume that we know exactly the regimes configuration and the type of model in each regime (OU model).

**Model evaluation**. In the next part of the simulation study, we wanted to assess the model selection aspect of our method. Using the same phylogenetic tree for the parameter estimation study, we defined three different MGPMs with increasing complexity, $\mathrm{M}_1$, $\mathrm{M}_2$, and $\mathrm{M}_3$. $\mathrm{M}_1$ contains only one regime, called *Ancestral*. In $\mathrm{M}_2$, there are two OU regimes called *Ancestral* and *R1*, while $\mathrm{M}_3$ contains four OU regimes, which are *Ancestral*, *R1*, *R2*, and *R3*. All MGPMs start from ancestral value $X_0$. The complete list of true parameters for each MGPM and their configuration on the tree can be seen in Table 1 and Fig. 3.

For each MGPM $\mathrm{M}_1$, $\mathrm{M}_2$, and $\mathrm{M}_3$, we simulated 100 sets of observations at the tips. Then, for each set of simulated observations, we assumed the configurations and models in $\mathrm{M}_1$, $\mathrm{M}_2$, and $\mathrm{M}_3$ (without the true parameters) as the three different models of evolution, and infer the posterior distributions of the parameters, given the set of observations

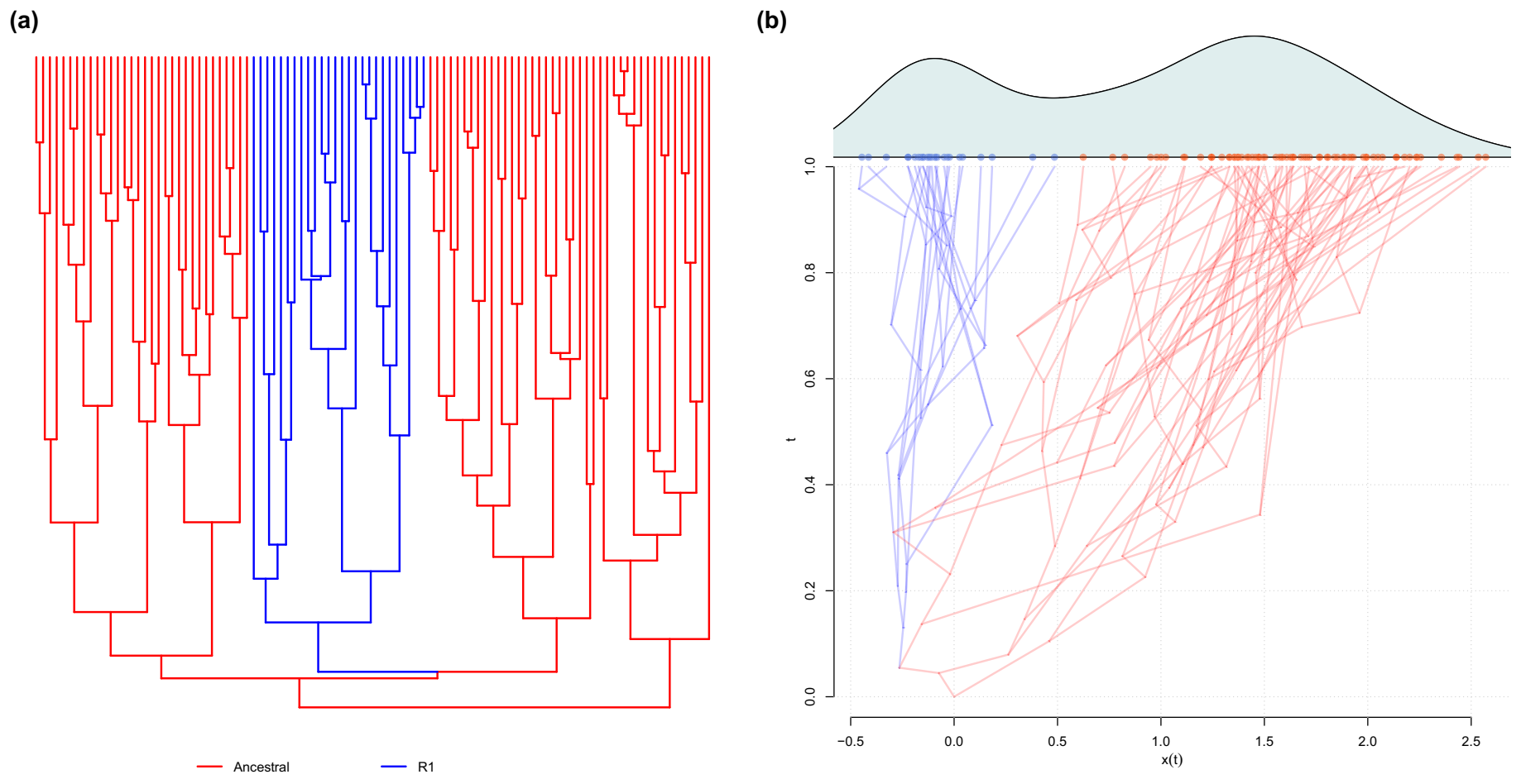

**Fig. 2** **a** Regimes configuration on the phylogenetic tree. **b** An illustration of the process to simulate observations at the tips by simulating values in the internal nodes of the tree

**Table 1** List of true parameters in all regimes for $\mathrm{M}_1$, $\mathrm{M}_2$, and $\mathrm{M}_3$. The parameter for ancestral value at the root, $X_0$, is set to the *Global* regime as it is shared among all other regimes

| MGPM | Regime | Model | True parameters |
|---|---|---|---|
| $\mathrm{M}_1$ | Global | – | $X_0 = 0$ |
| | Ancestral | OU | $\alpha = 2, \theta = 2, \sigma = 1$ |
| $\mathrm{M}_2$ | Global | – | $X_0 = 0$ |
| | Ancestral | OU | $\alpha = 2, \theta = 2, \sigma = 2$ |
| | R1 | OU | $\alpha = 1, \theta = -2, \sigma = 1$ |
| $\mathrm{M}_3$ | Global | – | $X_0 = 0$ |
| | Ancestral | OU | $\alpha = 3, \theta = 2, \sigma = 2$ |
| | R1 | OU | $\alpha = 2, \theta = 1, \sigma = 1$ |
| | R2 | OU | $\alpha = 2, \theta = -1, \sigma = 1$ |
| | R3 | OU | $\alpha = 3, \theta = -2, \sigma = 2$ |

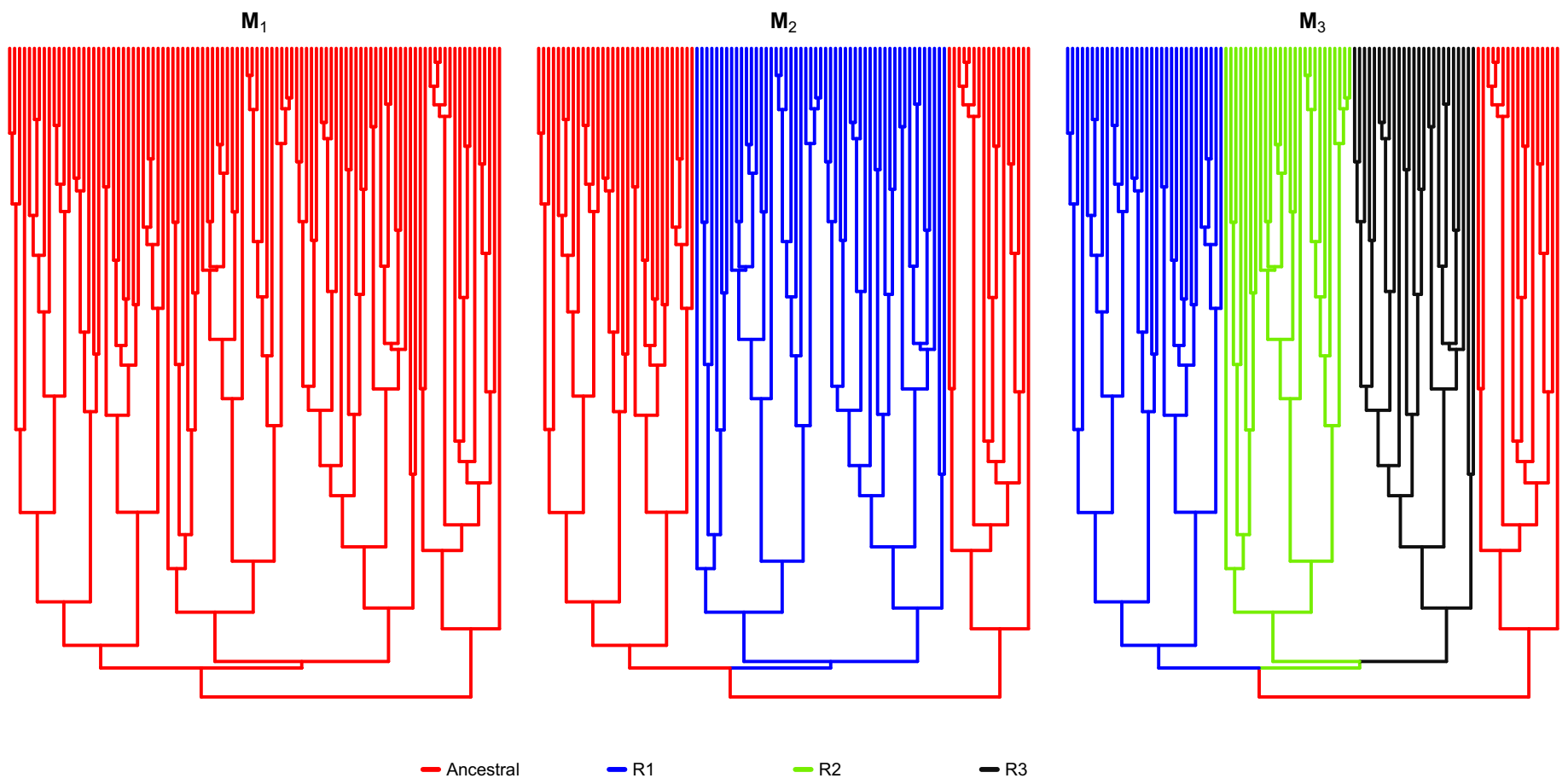

**Fig. 3** Configurations of evolutionary regimes for the MGPM $\mathrm{M}_1$, $\mathrm{M}_2$, and $\mathrm{M}_3$. For $\mathrm{M}_1$, there is only a single, global regime that is the same as the root node. The tree is split into two regimes for $\mathrm{M}_2$, and four regimes for $\mathrm{M}_3$

and the assumed model. In particular, let $\mathrm{x}_i^{(j)}$ be the *j*-th simulated observations from the MGPM $\mathrm{M}_i$. Then, we inferred the posterior distributions $p(\boldsymbol{\Theta}_k|\mathrm{x}_i^{(j)}, \mathrm{M}_k)$, for $k = 1, 2, 3$, where $\boldsymbol{\Theta}_k$ is the set of parameters of the assumed model $\mathrm{M}_k$. From each posterior inference, we computed the posterior predictive loss score, and compared them among different scenarios.

For the priors, we set Half-$\mathcal{N}(\sigma = 5)$ for all positive parameters ($\alpha's$ and $\theta$'s). For the unconstrained parameters ($X_0$'s and $\theta$'s) we assumed a $\mathcal{N}(\mu = 0, \sigma = 5)$ distribution as the prior.

### *Real-world data*

We used our method to study the evolution of antler size across the phylogeny of deer (family Cervidae) [5]. We chose posterior skull length (PSL), which we transformed into log-PSL values, as the proxy trait for the antler size and hypothesized three different scenarios describing the evolution of the antlers. First, we assume that the antler size evolved under the same process regardless of their position in the phylogenetic tree. Second, we assume that the antlers of the old-world and new-world deer species evolved under different processes; hence, the tree is split into two regimes. Third, we assume that the antler types, palmated (Palm), main beamed (MB), and bifurcated, have different evolutionary processes, and model the most common type, MB as the ancestral regime while mapping the regimes for palmated and bifurcated types onto terminals or clades where they are observed as tip data unequivocally. We proceed by denoting the MGPM for the first, second, and third scenarios as $\mathrm{M}_a$, $\mathrm{M}_b$, and $\mathrm{M}_c$, respectively. The regimes configurations of $\mathrm{M}_a$, $\mathrm{M}_b$, and $\mathrm{M}_c$ on the tree and the observed data at the tips can be seen in Fig. 4.

For all regimes, we assumed an OU process with parameters $\alpha$, $\theta$, and $\sigma$, unique to that regime. We further assumed that $X_0$ is shared between all regimes.

**Prior distribution**. We constructed weakly informative prior distributions for the parameters $X_0$, $\alpha$, $\theta$, and $\sigma$. We want a large portion of our prior distribution to cover realistic log-PSL values, but not specific enough that it would give very low probabilities to plausible log-PSL values.

For $X_0$ and $\theta$, we set a normal distribution with a 10% quantile at log(1 cm) = 0 and a 90% quantile at log(50 cm) $\approx$ 3.91202. Using these two points, we obtained a normal distribution $\mathcal{N}(\mu = 1.9560, \sigma^2 = 2.3295)$.

The parameter $\alpha$ determines the speed of convergence to $\theta$ from $X_0$. However, we can also relate it to the phylogenetic half-life [32],

$$t_{1/2} = \frac{\ln(2)}{\alpha}. \tag{32}$$

The interpretation is that an OU process starting from $X_0$ will reach half the distance to $\theta$ in $t_{1/2}$ units of time. We can deduce that a large value of $\alpha$ implies a shorter half-life, and vice versa.

Using the information above, we designed a prior based on a half-normal distribution with a scale parameter $\sigma$ so that the 95% quantile coincides with a value $\alpha^* = \ln(2)/t^*_{1/2}$ that we consider really high but still plausible. We assumed that the half-life could be as short as 1% of the height of the tree ($t_H$). This translates to $t^*_{1/2} = 0.1402$ my, and consequently $\alpha^* = 4.9410 \text{ my}^{-1}$. Therefore, we obtained Half-$\mathcal{N}(\sigma = 2.5210)$ as the prior distribution for $\alpha$, where the 95% quantile of this distribution is 4.9410.

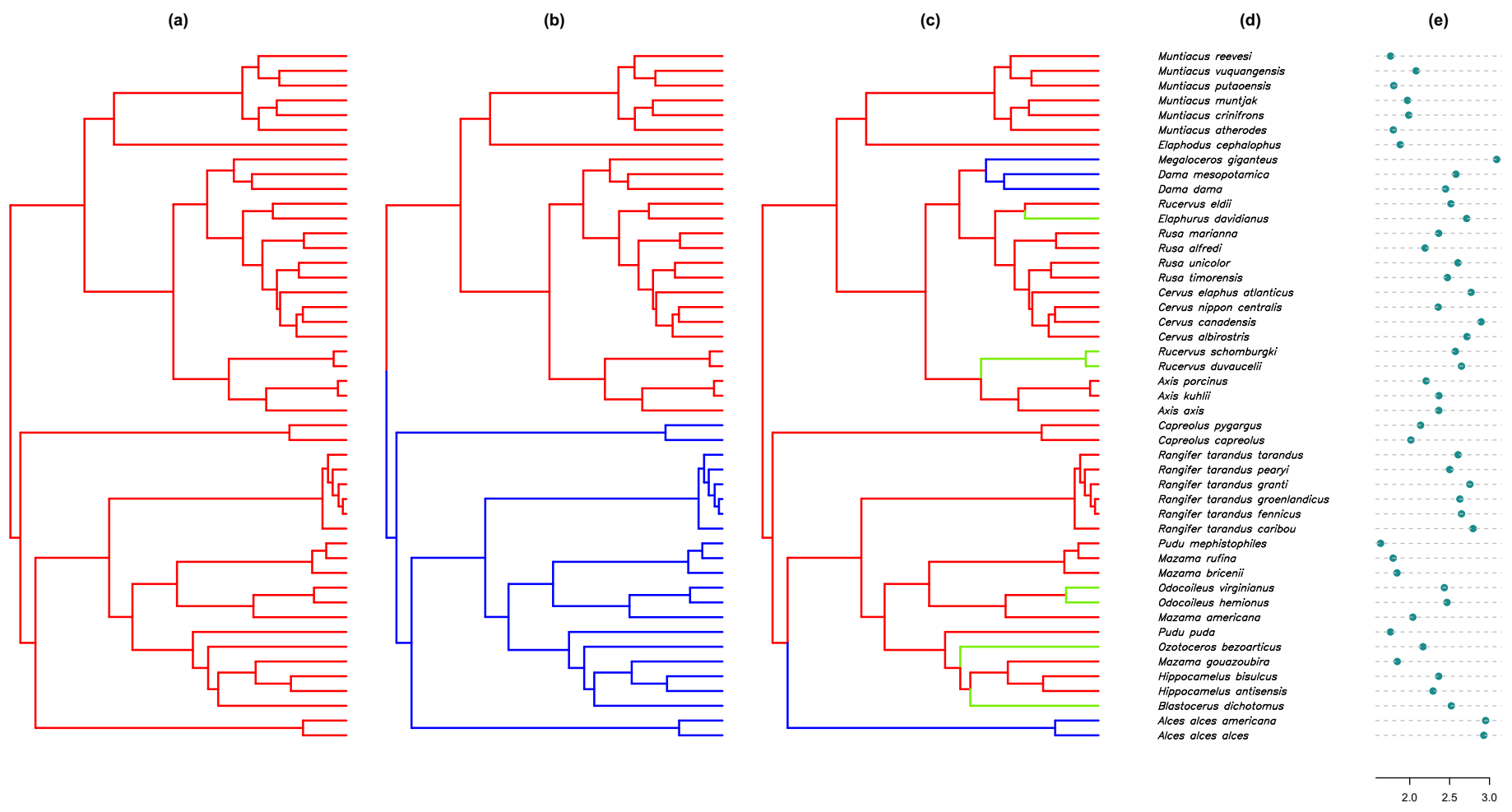

**Fig. 4** Different scenarios of regimes configuration on the deer phylogeny. In (**a**), there is only one regime for the whole tree. In (**b**), the red regime denotes the old-wold deer while the blue regime denotes the new-world deer. In the third scenario, **c** the tree is split into three regimes based on the antler shape: main beamed (MB, red), palmated (Palm, blue), and bifurcated (green). Subplot **(d)** lists the taxa names found in the tree, while subplot **e** shows the value of log of posterior skull length (PSL) for each taxon

The parameter $\sigma$ determines the strength of random fluctuations throughout the process of evolution. Depending on the type of process (BM or OU), $\sigma$ contributes to the variance of a trait at a given time $t$ as shown by equations (11) and (14). We used a similar procedure as for the $\alpha$ parameter, i.e., we used a half-normal distribution where the 95% quantile of this distribution is set to a value that is high enough but still realistic, denoted $\sigma^*$.

To set the value of $\sigma^*$, we assumed that the trait at the tip, $x_i(t_H)$, is a normal distribution with a variance twice as large as the prior distribution for $X_0$ and $\theta$. Furthermore, we assumed that $x_i(t_H)$ could have evolved from a BM process, which has a greater variance than an OU process. In other words,

$$x_i(t_H)|X_0 \sim \mathcal{N}(\mu = X_0, \sigma^2 = 4.6591) \tag{33}$$

From the distribution above, we obtained $\sigma^*$ through Equation (11):

$$\sigma^* = \sqrt{\frac{V_{t_H,BM,\sigma^*}}{t_H}} = \sqrt{\frac{4.6591}{14.02854}} = 0.5763. \tag{34}$$

Therefore, we set the half-normal distribution so that its 95% quantile is equal to $\sigma^* = 0.5763$, and obtained the Half-$\mathcal{N}(\sigma = 0.2940)$.

We applied the same prior distribution for the same type of parameter across all regimes. The complete list of parameters and their prior distributions for each MGPM is shown in Table 2. We compared the fit of each scenario to the observed data, measured by the posterior predictive loss score.

## Results

### Simulation study

**Parameter estimation**. Figure 5a shows the box plots of median of the marginal posterior distributions and the box plots of the maximum likelihood estimates, compared to the true values of the parameters. The medians of the marginal posterior distributions for $\theta_1$ and $\theta_2$ are the closest to the true values, followed by $\sigma_1$ and $\sigma_2$. Although the medians of $X_0$ are mostly around the true values, the medians of the rate parameters $\alpha_1$ and $\alpha_2$ are larger than the true values.

**Table 2** List of models on the regimes of $\mathrm{M}_a$, $\mathrm{M}_b$, and $\mathrm{M}_c$, with their parameters and priors. The parameter for ancestral value at the root, $X_0$, is set to the *Global* regime as it is shared among all other regimes

| MGPM | Regime | Model | Parameters and priors |
|---|---|---|---|
| $\mathrm{M}_a$ | Global | – | $X_0 \sim \mathcal{N}(\mu = 1.9560, \sigma^2 = 2.3295)$ |
| | Ancestral | OU | $\alpha \sim$ Half-$\mathcal{N}(\sigma = 2.5210)$, $\theta \sim \mathcal{N}(\mu = 1.9560, \sigma^2 = 2.3295)$, $\sigma \sim$ Half-$\mathcal{N}(\sigma = 0.2940)$ |
| $\mathrm{M}_b$ | Global | – | $X_0 \sim \mathcal{N}(\mu = 1.9560, \sigma^2 = 2.3295)$ |
| | Old-world | OU | $\alpha_1 \sim$ Half-$\mathcal{N}(\sigma = 2.5210)$, $\theta_1 \sim \mathcal{N}(\mu = 1.9560, \sigma^2 = 2.3295)$, $\sigma_1 \sim$ Half-$\mathcal{N}(\sigma = 0.2940)$ |
| | New-world | OU | $\alpha_2 \sim$ Half-$\mathcal{N}(\sigma = 2.5210)$, $\theta_2 \sim \mathcal{N}(\mu = 1.9560, \sigma^2 = 2.3295)$, $\sigma_2 \sim$ Half-$\mathcal{N}(\sigma = 0.2940)$ |
| $\mathrm{M}_c$ | Global | – | $X_0 \sim \mathcal{N}(\mu = 1.9560, \sigma^2 = 2.3295)$ |
| | MB | OU | $\alpha_1 \sim$ Half-$\mathcal{N}(\sigma = 2.5210)$, $\theta_1 \sim \mathcal{N}(\mu = 1.9560, \sigma^2 = 2.3295)$, $\sigma_1 \sim$ Half-$\mathcal{N}(\sigma = 0.2940)$ |
| | Palm | OU | $\alpha_2 \sim$ Half-$\mathcal{N}(\sigma = 2.5210)$, $\theta_2 \sim \mathcal{N}(\mu = 1.9560, \sigma^2 = 2.3295)$, $\sigma_2 \sim$ Half-$\mathcal{N}(\sigma = 0.2940)$ |
| | Bifurcated | OU | $\alpha_3 \sim$ Half-$\mathcal{N}(\sigma = 2.5210)$, $\theta_3 \sim \mathcal{N}(\mu = 1.9560, \sigma^2 = 2.3295)$, $\sigma_3 \sim$ Half-$\mathcal{N}(\sigma = 0.2940)$ |

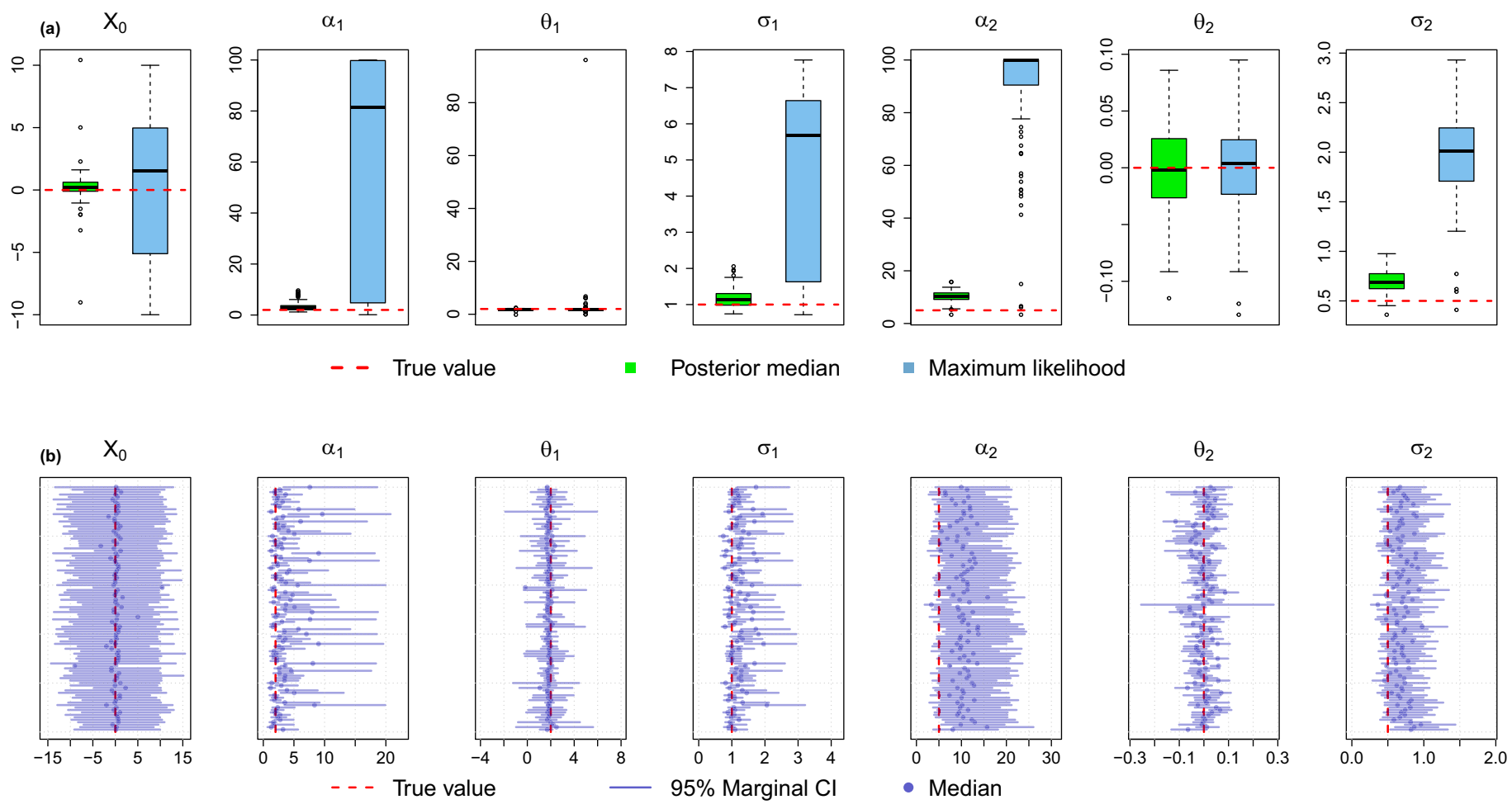

**Fig. 5** **a** Box plots of 100 posterior medians (light green) and maximum likelihood estimates (light blue) compared to the true parameter values (red dashed lines). **b** The 95% equal-tailed credible intervals of the marginal posterior distribution. There are 100 intervals shown for each parameter, along with the medians

**Table 3** Summary of the 100 95% credible intervals (± standard deviation) obtained from the posterior inferences

| Parameter | True value | Mean lower bound of 95% CI | Mean upper bound of 95% CI | Mean width |
|---|---|---|---|---|
| $X_0$ | 0 | −10.2469 ± 2.1264 | 10.6607 ± 1.9772 | 20.9076 ± 3.5001 |
| $\alpha_1$ | 2 | 1.9528 ± 0.8494 | 6.9801 ± 5.4140 | 5.0274 ± 4.6572 |
| $\theta_1$ | 2 | 0.8792 ± 0.7854 | 2.8289 ± 0.9175 | 1.9498 ± 1.6700 |
| $\sigma_1$ | 1 | 0.9506 ± 0.1567 | 1.6713 ± 0.5780 | 0.7207 ± 0.4422 |
| $\alpha_2$ | 5 | 4.6857 ± 1.1737 | 20.1030 ± 3.0438 | 15.4174 ± 2.5894 |
| $\theta_2$ | 0 | −0.0579 ± 0.0477 | 0.0527 ± 0.0433 | 0.1106 ± 0.0535 |
| $\sigma_2$ | 0.5 | 0.4609 ± 0.0819 | 1.0396 ± 0.1736 | 0.5787 ± 0.1176 |

We can also see the posterior distribution of the parameters through their 95% equal-tailed credible intervals, which is shown in Fig. 5b. Although the medians of the marginal posterior distribution for some parameters are higher than the true values, i.e., $\alpha_1$, $\alpha_2$, $\sigma_1$, and $\sigma_2$, most of the time the true values are still included in the credible intervals. In Table 3, we can see the summary of the 100 95% credible intervals for the parameters. We can observe that the true parameter values are within the means of the lower and upper bounds of the intervals.

Compared to the maximum likelihood estimates, the posterior medians are in general closer to the true values. In some cases where the medians are larger than the true values, i.e., $\alpha_1$, $\alpha_2$, $\sigma_1$, and $\sigma_2$, the maximum likelihood estimates are much higher. This is due to the fact that the prior distributions, even though they are diffuse, put lower probabilities on unrealistic parameter values, e.g., high $\alpha$ or $\sigma$. This could result in posterior distributions that are more confined into more realistic parameter values, and this might be a remedy to one of the problems of maximum likelihood based inference for PCMs, that is, large estimates of $\alpha$ and $\sigma$ (see, e.g., [33]). Moreover, maximum likelihood

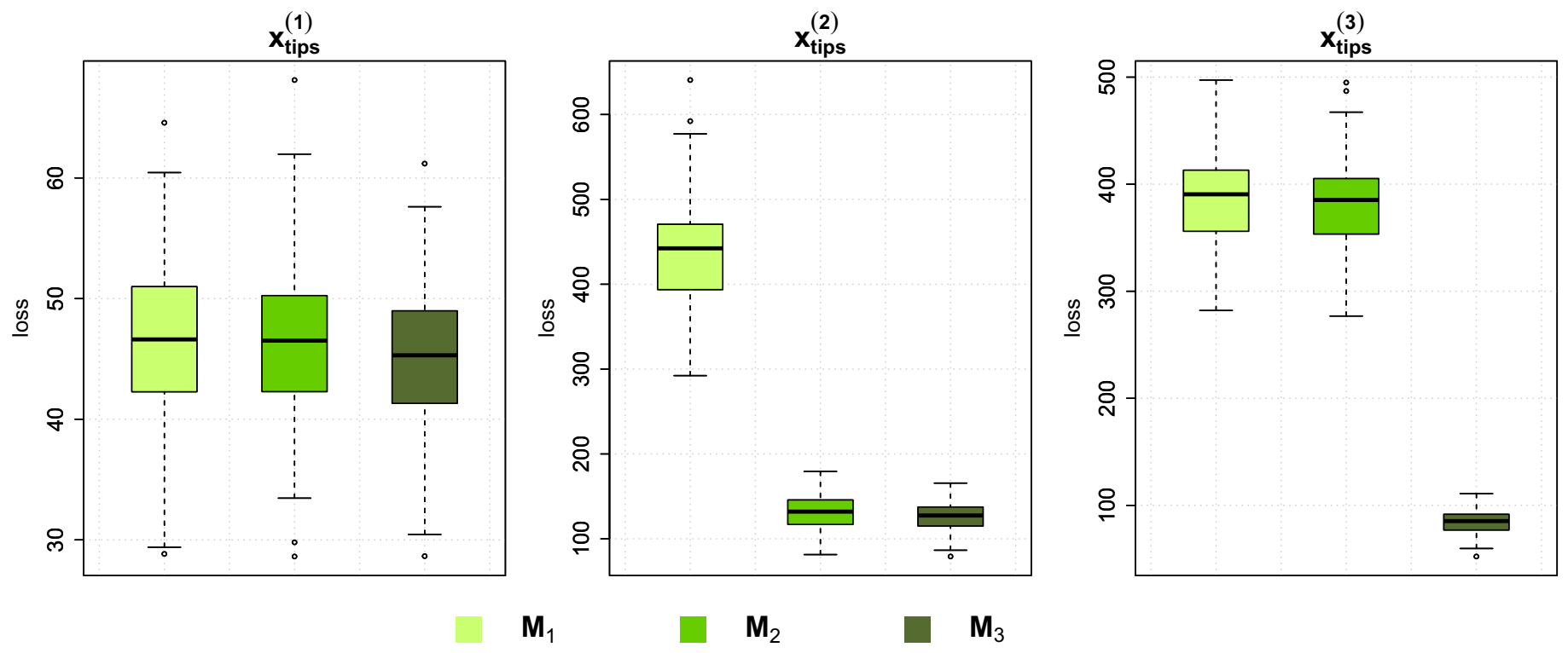

**Fig. 6** Box plots of posterior predictive loss scores. The left-hand-side shows the box plots for posterior predictive loss scores from MGPMs $\mathrm{M}_1$, $\mathrm{M}_2$, and $\mathrm{M}_3$, given that the observations are simulated from $\mathrm{M}_1$. Similarly, the center and the right-hand-side plot shows posterior predictive loss scores when the observations are simulated from $\mathrm{M}_2$ and $\mathrm{M}_3$, respectively

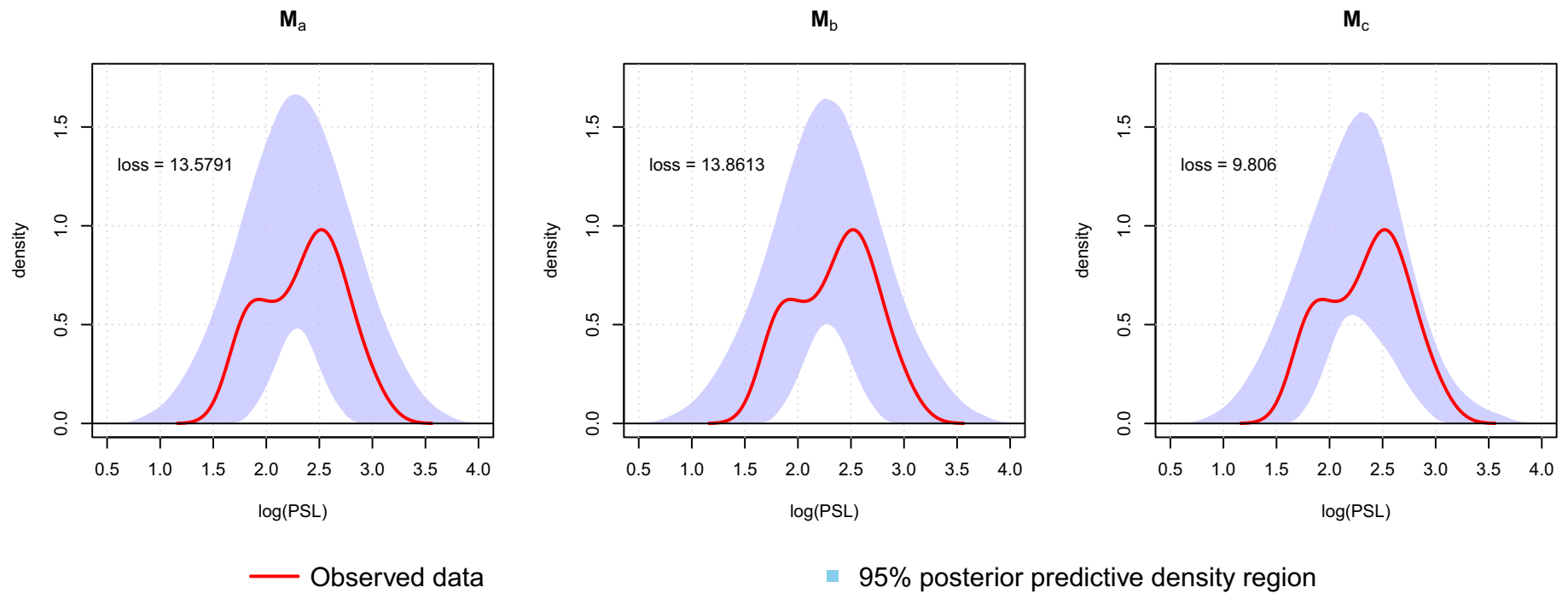

**Fig. 7** From left to right: Density of the observed data compared to the density of the simulated data from the posterior predictive distribution assuming $\mathrm{M}_a$, $\mathrm{M}_b$, or $\mathrm{M}_c$ as the model of evolution

estimates are found by performing optimization routines on the likelihood space, which could be multimodal or even flat. Hence, the optimization algorithm could easily become stuck at a local maximum.

**Model evaluation**. The posterior predictive loss values of the MGPMs $\mathrm{M}_1$, $\mathrm{M}_2$, and $\mathrm{M}_3$, given the observed data simulated by each of them, can be seen in Fig. 6. In the case where observations are simulated from $\mathrm{M}_1$, the three models are comparable in terms of their posterior predictive loss scores. Once we observed a more complicated set of observations simulated from $\mathrm{M}_2$, we can see that the models $\mathrm{M}_2$ and $\mathrm{M}_3$ have lower scores than $\mathrm{M}_1$. Finally, $\mathrm{M}_3$ has the lowest scores when the observations are generated by $\mathrm{M}_3$.

## Real-world data

The comparison of the posterior predictive distributions for MGPMs $\mathrm{M}_a$, $\mathrm{M}_b$, and $\mathrm{M}_c$ can be seen in Fig. 7, which shows the 95% credible region of kernel density estimates from the simulated data from the posterior predictive distribution of each MGPM. The MGPM $\mathrm{M}_b$, in which the regimes are based on old-world and new-world deer, has a higher posterior predictive loss score than $\mathrm{M}_a$, which assumes the same OU process throughout the tree. This suggests that the regimes configuration based on taxonomic definitions of old-world and new-world deer does not result in a model that can better explain the observed data than the single-OU model. On the other hand, the MGPM $\mathrm{M}_c$, which involves three types of regimes, has a lower score than the other two MGPMs. This is also indicated by the credible region of the simulated data from the posterior predictive distribution of $\mathrm{M}_c$, which is more narrow, suggesting that the simulated data from the posterior predictive distribution are closer to the observed data.

To take a closer look at the resulting posterior predictive distributions of the distinct MGPMs, we can analyze the simulated posterior predictive distributions at the tips. From Equation (28), we know that the posterior predictive density, $p(\tilde{\mathrm{x}}|\mathrm{x})$ is calculated by integrating the likelihood function $p(\tilde{\mathrm{x}}|\boldsymbol{\Theta})$ with respect to the posterior distribution $p(\boldsymbol{\Theta}|\mathrm{x})$. Furthermore, we know that the likelihood $p(\tilde{\mathrm{x}}|\boldsymbol{\Theta})$ is a multivariate normal. Hence, we can estimate the posterior predictive density by approximating the integral in Equation (28), which is done by calculating the following formula:

$$\hat{p}(\tilde{\mathrm{x}}|\mathrm{x}) := \frac{1}{S}\sum_{i=1}^{S} p(\tilde{\mathrm{x}}|\boldsymbol{\Theta}_i), \tag{35}$$

where $\boldsymbol{\Theta}_i, \ldots, \boldsymbol{\Theta}_S$ are the $S$ number of draws from the posterior distribution.

Figure 8 shows the estimated posterior predictive densities at the tips using $S = 1000$ on $\mathrm{M}_a$, $\mathrm{M}_b$, and $\mathrm{M}_c$. The posterior predictive densities of $\mathrm{M}_b$ are similar to $\mathrm{M}_a$, even though the former has two regimes. This explains the posterior predictive loss value of $\mathrm{M}_b$, which is similar to $\mathrm{M}_a$. We can also see that $\mathrm{M}_c$ has posterior predictive densities that are different depending on their regimes. If we look at the locations of the posterior predictive distributions, the species in the Palm regime tend to have a higher mean, followed by the species in the Bifurcated and MB regimes. The posterior predictive distribution in the $\mathrm{M}_c$ case explains its lower posterior predictive loss values, where the sum square difference between the posterior predictive means and the observed data is lower due to the regimes that produce more flexible posterior predictive distributions.

To make sense of the distinct distributions on different regimes in the posterior predictive distribution in $\mathrm{M}_c$, we can look at Fig. 9 which shows the simulated marginal posterior density of the parameters of $\mathrm{M}_c$, and Table 4 which presents the mean, median, standard deviation, and 95% equal-tail credible interval of the marginal posterior distribution.

We can see that the marginal posterior densities of the parameters from different regimes are easily distinguishable. In the marginal posterior densities for $\theta$, we can clearly see that the modes do not overlap, suggesting that each regime has its own stabilizing optimum. However, the marginal posterior distribution of $\theta$ for the MB regime (red) is wider than the others, with 95% of its mass between 1.4517 and 2.9394. This can be related to the marginal posterior density of $\alpha$, where the MB regime is concentrated around values close to zero, with a mean close to 0.1. Lower values of $\alpha$ indicate that the traits under the MB regime behave more closely to a BM process, where there is no particular direction of evolution.

This also explains the posterior predictive distributions in Fig. 8, where the species in the Palm regime have a higher mean, followed by the Bifurcated and MB regimes. Since $\theta$ is the long-term mean of the OU process (equations (7), (14)), a posterior distribution with a higher mean for $\theta$ translates to a posterior predictive distribution with a higher mean.

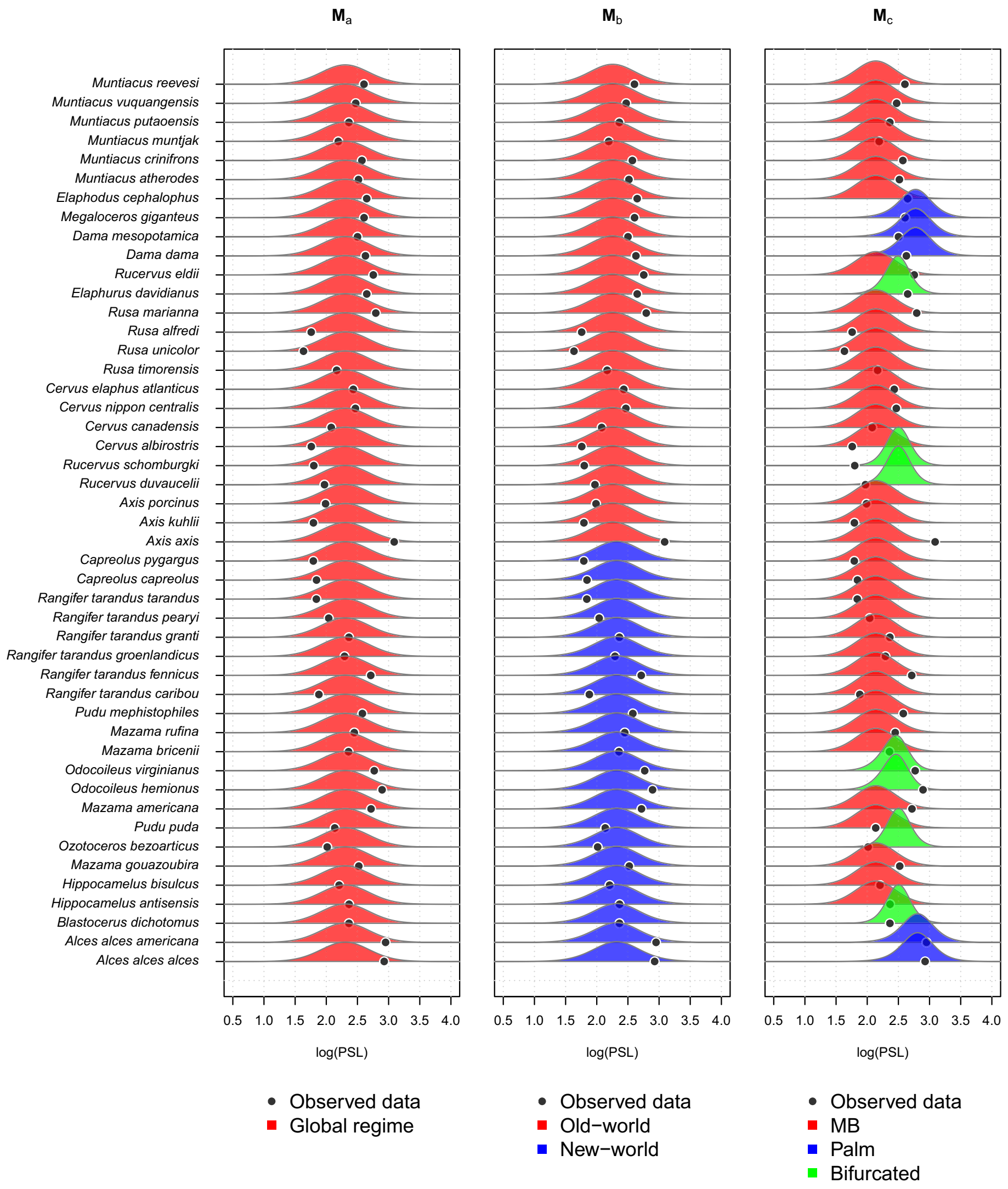

**Fig. 8** From left to right: Estimated posterior predictive density curves at the tips, assuming $\mathrm{M}_a$, $\mathrm{M}_b$, or $\mathrm{M}_c$

Regarding the marginal posterior distribution of $\sigma$, Fig. 9 shows that for the MB regime, the marginal posterior distribution is approximately 0.1442, with a 95% credible interval between 0.1137 and 0.1837. This is significantly narrower than the other two regimes, where the Palm regime has a 95% credible interval between 0.1683 and 0.6460, and the Bifurcated regime has a 95% credible interval between 0.1501 and 0.5282. This could indicate that there are more random fluctuations in the Palm and Bifurcated regimes, which have shorter periods compared to MB.

We can also notice how the shapes of the marginal posterior distributions differ from the prior of the parameters. The widths of the marginal posteriors of $\theta$'s decrease significantly compared to their prior, indicating that the likelihood surfaces are more informative, i.e., more concentrated for $\theta$'s. This phenomenon is also observed for the parameters $\alpha$ and $\sigma$ for the MB regime, where both are highly concentrated compared to the other regimes. The marginal posterior for $X_0$ does not change much from its prior, suggesting that the likelihood surface on $X_0$ is flatter.

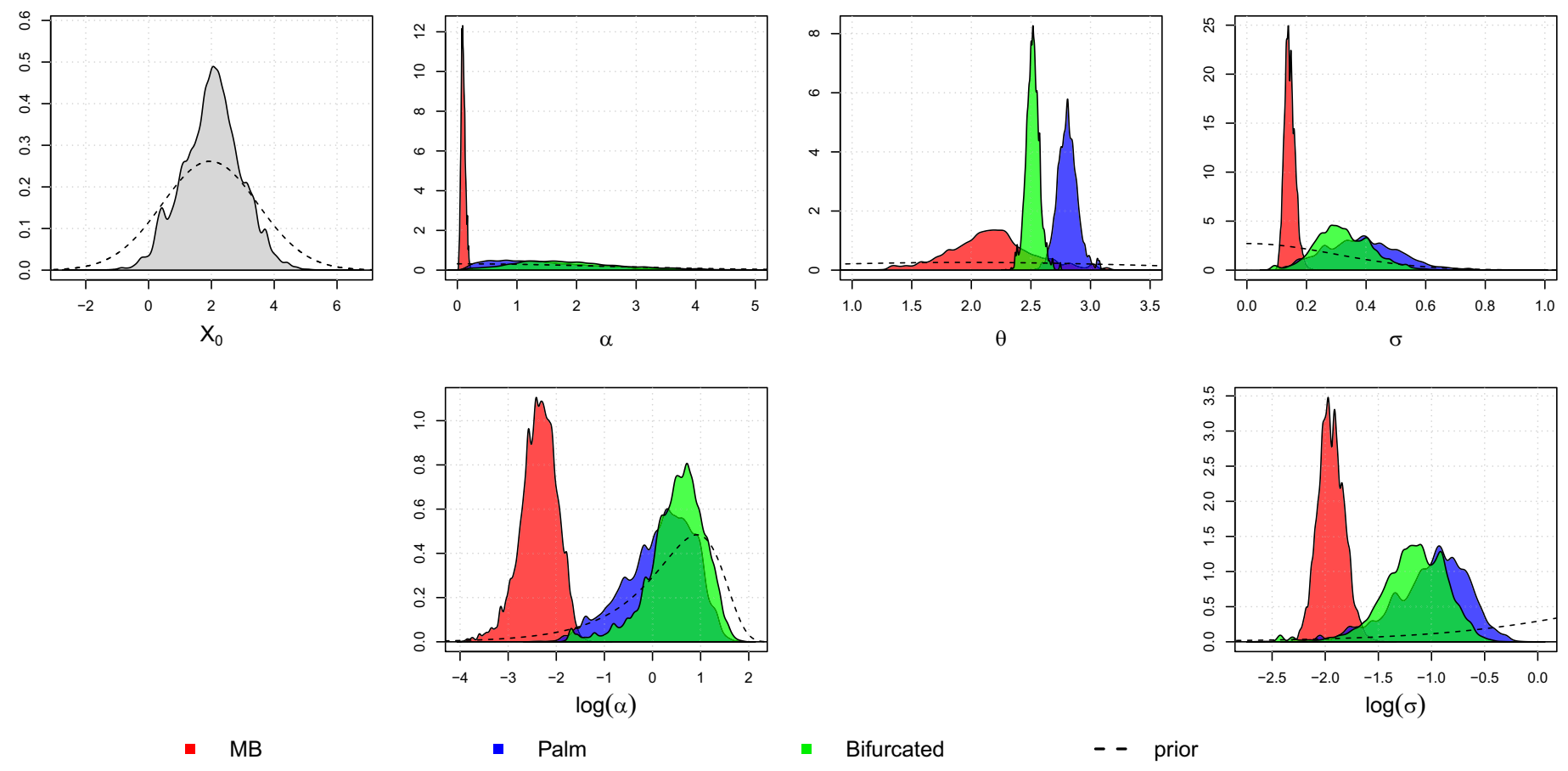

**Fig. 9** Simulated marginal posterior density of parameters of MGPM $\mathrm{M}_c$. The parameters are color-coded based on the regime they are in: red for MB, blue for Palm, and green for Bifurcated. The parameter $X_0$ is shared between all regimes. The prior densities are shown by the dashed lines. The top row shows marginal densities on the original space of the parameter, while the bottom row shows the marginal densities on the log scale for parameters $\alpha$ and $\sigma$

**Table 4** The mean, median, standard deviation, and 95% equal-tail credible interval of marginal posterior distribution of parameters of MGPM $\mathrm{M}_c$

| Regime | Parameter | Mean | Median | Std. dev. | 95% CI |
|---|---|---|---|---|---|
| – | $X_0$ | 2.0185 | 2.0445 | 0.9364 | (0.2104, 3.8204) |
| MB | $\alpha$ | 0.1002 | 0.0960 | 0.0360 | (0.0416, 0.1816) |
| | $\theta$ | 2.1666 | 2.1613 | 0.3468 | (1.4517, 2.9394) |
| | $\sigma$ | 0.1442 | 0.1421 | 0.0177 | (0.1137, 0.1837) |
| Palm | $\alpha$ | 1.4477 | 1.2905 | 0.8867 | (0.2425, 3.5964) |
| | $\theta$ | 2.8058 | 2.8045 | 0.0846 | (2.6438, 2.9903) |
| | $\sigma$ | 0.3869 | 0.3845 | 0.1236 | (0.1683, 0.6460) |
| Bifurcated | $\alpha$ | 1.9145 | 1.7679 | 0.9694 | (0.3706, 4.1251) |
| | $\theta$ | 2.5155 | 2.5151 | 0.0565 | (2.3992, 2.6325) |
| | $\sigma$ | 0.3251 | 0.3168 | 0.0945 | (0.1501, 0.5282) |

Furthermore, we can observe the marginal posterior densities of parameters $\alpha$ and $\sigma$ on the transformed (log) scale in the bottom row of Fig. 9. We observe significant changes in the marginal densities of $\alpha$'s before and after the log transformation. For the MB regime, the marginal posterior density for log-$\alpha$ is concentrated around more negative values. This transforms into a density that is highly concentrated around a value close to zero in the original space. For the Palm and Bifurcated regimes, the log-$\alpha$ densities are stretched because they cover very wide values of log-$\alpha$. On the other hand, the densities of log-$\sigma$ retain relatively similar shapes when transformed to the original space.

### Prior sensitivity analysis

In 2.4, we used diffuse priors to demonstrate how the Bayesian inference method works without specific prior preferences. To further analyze the effects of prior choices on parameter inference and model evaluation results, we reran the simulation study with different prior configurations.

For parameter inference, we implemented two new prior configurations that are progressively more informative than the original prior used in 2.4. We denote prior 2 and prior 3 as the new prior configurations, while prior 1 is the original prior.

The prior distributions of the parameters under prior 2 are

$$
\begin{aligned}
X_0 &\sim \mathcal{N}(\mu = 0, \sigma = 5),\\
\alpha_1 &\sim \text{Half-}\mathcal{N}(\sigma = 5),\\
\theta_1 &\sim \mathcal{N}(\mu = 2, \sigma = 5),\\
\sigma_1 &\sim \text{Half-}\mathcal{N}(\sigma = 5),\\
\alpha_2 &\sim \text{Half-}\mathcal{N}(\sigma = 5),\\
\theta_2 &\sim \mathcal{N}(\mu = 0, \sigma = 5),\\
\sigma_2 &\sim \text{Half-}\mathcal{N}(\sigma = 5),
\end{aligned}
$$

while prior 3 can be described as follows.

$$
\begin{aligned}
X_0 &\sim \mathcal{N}(\mu = 0, \sigma = 2),\\
\alpha_1 &\sim \text{Half-}\mathcal{N}(\sigma = 1),\\
\theta_1 &\sim \mathcal{N}(\mu = 2, \sigma = 2),\\
\sigma_1 &\sim \text{Half-}\mathcal{N}(\sigma = 1),\\
\alpha_2 &\sim \text{Half-}\mathcal{N}(\sigma = 5),\\
\theta_2 &\sim \mathcal{N}(\mu = 0, \sigma = 2),\\
\sigma_2 &\sim \text{Half-}\mathcal{N}(\sigma = 1).
\end{aligned}
$$

We conducted another study of model evaluation with a more informative configuration for the priors than in 2.4. The list of new prior distributions of the parameters of models $\mathrm{M}_1$, $\mathrm{M}_2$, $\mathrm{M}_3$, can be found in Table 5.

The parameter inference and model evaluation results are shown in Fig. 10. As expected, a more informative prior distribution leads to a posterior distribution that is closer to the true parameters. At the same time, we can also observe that the variances of the posterior medians are decreased for models with a more informative prior distribution. The results of the model evaluation study with a narrower prior show a pattern similar to that in Fig. 6, where, in a more complex dataset, a simpler model has a worse (higher) posterior predictive loss score.

#### Posterior predictive loss analysis

The results of the simulation study show that the more complex model can achieve a posterior predictive loss score similar to that of the simpler model. This could be interpreted as all models explaining the generative process of the observations equally well. In the context of Equation (30), a low posterior predictive loss score could mean that the mean of the posterior predictive distribution is close to the observed data, the variance

**Table 5** List of prior distributions in all regimes for $\mathrm{M}_1$, $\mathrm{M}_2$, and $\mathrm{M}_3$

| MGPM | Regime | Model | Prior distributions |
|---|---|---|---|
| $\mathrm{M}_1$ | Global | – | $X_0 \sim \mathcal{N}(\mu = 0, \sigma = 2)$ |
| | Ancestral | OU | $\alpha \sim \text{Half-}\mathcal{N}(\sigma = 2), \theta \sim \mathcal{N}(\mu = 2, \sigma = 2), \sigma \sim \text{Half-}\mathcal{N}(\sigma = 1)$ |
| $\mathrm{M}_2$ | Global | – | $X_0 \sim \mathcal{N}(\mu = 0, \sigma = 2)$ |
| | Ancestral | OU | $\alpha \sim \text{Half-}\mathcal{N}(\sigma = 2), \theta \sim \mathcal{N}(\mu = 2, \sigma = 2), \sigma \sim \text{Half-}\mathcal{N}(\sigma = 2)$ |
| | R1 | OU | $\alpha \sim \text{Half-}\mathcal{N}(\sigma = 1), \theta \sim \mathcal{N}(\mu = -2, \sigma = 2), \sigma \sim \text{Half-}\mathcal{N}(\sigma = 1)$ |
| $\mathrm{M}_3$ | Global | – | $X_0 \sim \mathcal{N}(\mu = 0, \sigma = 2)$ |
| | Ancestral | OU | $\alpha \sim \text{Half-}\mathcal{N}(\sigma = 3), \theta \sim \mathcal{N}(\mu = 2, \sigma = 2), \sigma \sim \text{Half-}\mathcal{N}(\sigma = 2)$ |
| | R1 | OU | $\alpha \sim \text{Half-}\mathcal{N}(\sigma = 2), \theta \sim \mathcal{N}(\mu = 1, \sigma = 2), \sigma \sim \text{Half-}\mathcal{N}(\sigma = 1)$ |
| | R2 | OU | $\alpha \sim \text{Half-}\mathcal{N}(\sigma = 2), \theta \sim \mathcal{N}(\mu = -1, \sigma = 2), \sigma \sim \text{Half-}\mathcal{N}(\sigma = 1)$ |
| | R3 | OU | $\alpha \sim \text{Half-}\mathcal{N}(\sigma = 3), \theta \sim \mathcal{N}(\mu = -2, \sigma = 2), \sigma \sim \text{Half-}\mathcal{N}(\sigma = 2)$ |

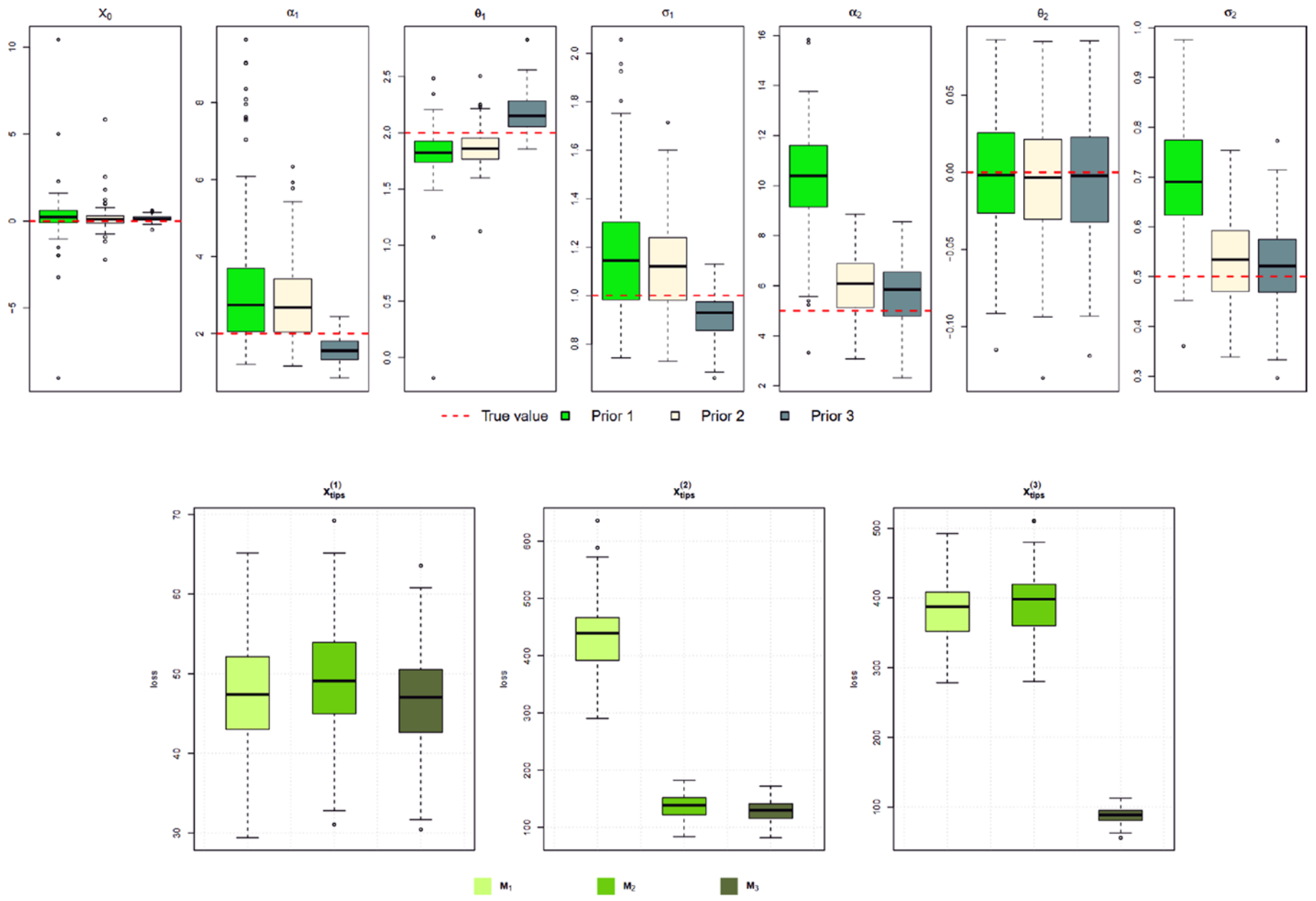

**Fig. 10** Top: Box plots of posterior median values from 100 parameter inference runs with different prior settings. Bottom: Box plots of posterior predictive loss scores of different MGPMs that are fitted to different data

of the posterior predictive distribution is small, or both. A more complex model could easily produce a posterior predictive distribution with a mean close to the observed data. However, there could be cases where the variance term is low, even though the model is more complex. As a simple example, consider a mixture of two normal distributions,

$$\frac{1}{2}\mathcal{N}(\mu_1, \sigma_1^2) + \frac{1}{2}\mathcal{N}(\mu_2, \sigma_2^2), \tag{36}$$

which is a more complex model than a single normal distribution $\mathcal{N}(\mu, \sigma^2)$, but could have a lower variance if $\sigma_1^2$ and $\sigma_2^2$ are small enough.

To alleviate this, we can multiply the variance term in Equation (30) by a positive scalar $\lambda \geq 1$, so that the modified estimated posterior predictive loss score is

$$\widehat{\text{post. pred. loss}} = \sum_{i=1}^{n}(x_i - \tilde{m}_i)^2 + \lambda \sum_{i=1}^{n} \tilde{s}_i^2. \tag{37}$$

The scalar $\lambda$ can be tied to a measure of the model's complexity. We take examples where $\lambda = p$ and $\lambda = \ln(p)$, where *p* is the number of parameters in the given MGPM, and we recalculated the posterior predictive loss scores for the model evaluation simulation study in 3.3. Figure 11 shows the scores of the modified posterior predictive loss. Notice that in both cases of $\lambda$, the true model that generates the data has, on average, the lowest loss score.

We also applied the modified posterior predictive loss function to the real-world data results in 2.4. In Table 6, we can see that using $\lambda = p$ results in a high penalty term for model complexity. Hence, $\mathrm{M}_a$, has the lowest posterior predictive loss score. Using $\lambda = \ln(p)$, we obtain a conclusion similar to the original analysis, where $\mathrm{M}_c$ has the lowest posterior predictive score, consequently followed by $\mathrm{M}_a$ and $\mathrm{M}_b$. In both cases, $\mathrm{M}_b$ always has the highest loss score, implying that it has a worse fit in terms of the sum-squared errors in Equation (37).

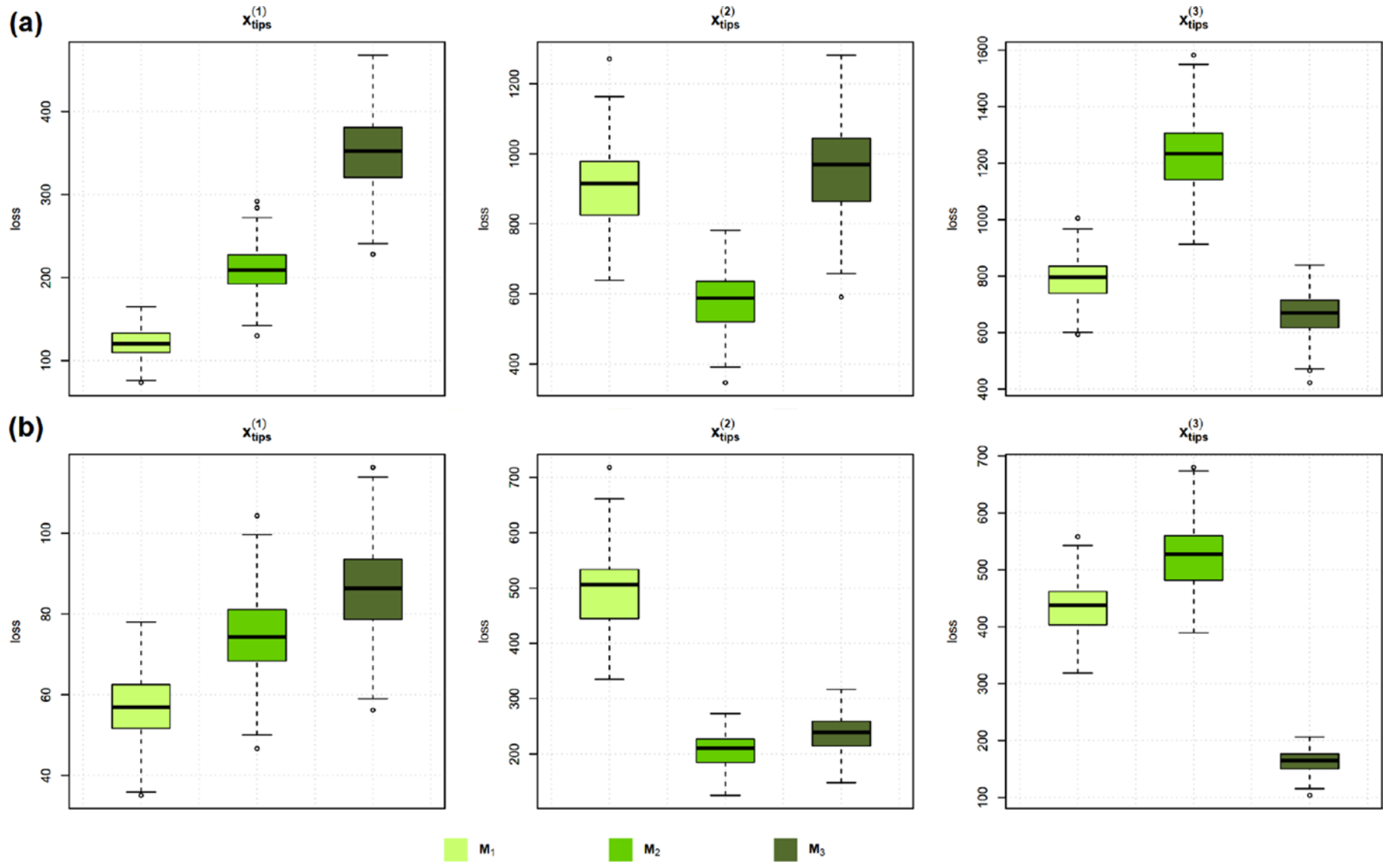

**Fig. 11** Box plots of the modified posterior predictive loss scores of different MGPMs that are fitted to different data. In (**a**), $\lambda = p$ is used and $\lambda = \ln(p)$ in (**b**)

**Table 6** Posterior predictive loss scores on the real-world application with $\lambda = p$ and $\lambda = \ln(p)$, where *p* is the number of parameters in the MGPM

| MGPM | PPL ($\lambda = p$) | PPL ($\lambda = \ln(p)$) |
|---|---|---|
| $\mathrm{M}_a$ ($p = 4$) | 34.3403 | 16.2524 |
| $\mathrm{M}_b$ ($p = 7$) | 55.8687 | 20.4838 |
| $\mathrm{M}_c$ ($p = 10$) | 48.8057 | 15.4505 |

## Discussion

Simulation studies showed that our method can properly infer the posterior distribution of MGPM parameters. The posterior distribution of the MGPM parameters inferred using our method is in general closer to the true parameters that generated the simulated data, compared to the maximum likelihood estimates. This is not exactly a fair comparison since we only compared point estimates of the posterior distribution with the maximum likelihood values. However, this shows that the PMC algorithm that we implemented to perform the Bayesian inference managed to capture the high posterior density area, which is mainly driven by the likelihood term since we used diffuse priors in the simulation study. Additionally, adding more information into the prior makes the posterior distribution closer to the true parameters.

The advantage of using a prior distribution is apparent in preventing inference to explore parts of the parameter space that result in unrealistic parameter values. In spite of that, defining a proper prior distribution can be challenging, especially when there is a lot of uncertainty around the parameters and when the observed data are often only available at the tips. Furthermore, problems with model identifiability persist, as OU models have inherent identifiability issues [16]. The prior distribution can only partially help to remedy the problem [17], but the inclusion of fossil data could further help, specifically to break the symmetry between different models or regimes [34].

Evidently, the posterior predictive loss is capable of quantifying whether the hypothesized MGPM agrees with the observations. In the simulation study, we observed that simple models have higher posterior predictive loss scores on a complex data, compared to models with more parameters. We further analyzed this using a modified posterior predictive loss with a parameter $\lambda$, which is associated with the complexity of the model. We found that incorporating an explicit term for model complexity into the penalizing term of the posterior predictive loss score helps alleviate overfitting of the more complex model to the simpler data. However, more studies are needed to analyze the effects of $\lambda$ in different scenarios. Other similar metrics based on the predictive capacity of the model may also be considered, such as widely applicable information criterion (WAIC, [35]) and cross-validation [36].

It should be noted that in the simulation study for parameter estimation, we always assumed that we know exactly the true regimes and their true models, both for the posterior and maximum likelihood setting (without the true parameters). This is done to emulate the condition where there is no model mismatch. Hence, the conclusions drawn from the experiment only concern the parameter inference capability. In the model evaluation study, we included one true MGPM among the three hypotheses, which is an ideal condition where we have the true model as one of our hypotheses. However, in the real-life scenario, we cannot know for sure if a true model is included among our proposed hypotheses or if it even exists. This scenario falls under the $\mathcal{M}$-completed or $\mathcal{M}$-open problem [37], where the latter assumes that the true model may be too difficult to formulate.

We illustrate how our method can be used to test several evolutionary hypotheses in a real world setting. We showed how an MGPM with multiple regimes does not necessarily result in a better fit to the observed data than using a single process on the whole tree. However, a regime configuration based on a more natural classification of traits, i.e., shape of the antlers, results in an MGPM that is more plausible in explaining the process that generated observed data.

We would like to underline that we do not write that we have found the "true" model, since it might not be within our considered hypotheses. What we are interested in is the comparison of multiple models that could explain the data-generating process. A better (lower) posterior predictive loss score could indicate that a model explains the data-generating process better, and hence captures some key properties in the data. However, one could easily construct such an overly complex model that excels in terms of posterior predictive loss score. In this sense, the posterior predictive loss score does not act as an absolute score but as an indicator of how much the model agrees with the observed data and each model under consideration should correspond to a carefully thought-through hypothesis so that the indicated model is interpretable (but this is essentially a warning common to any model selection procedure).

An obvious limitation of our method lies in the assumptions of fixed tree and fixed regimes. Since the tree itself is a result of inference from molecular sequences, it would make more sense to include the uncertainties regarding the tree itself by including the molecular data. This has been done previously in multiple studies (see, e.g., [18, 38, 39]). On top of that, while the method is built on `PCMBase` engine that utilizes fast likelihood calculation, the posterior inference is implemented in pure `R`, which leaves room for improvements in terms of speed.

## Conclusion

In this work, we implement a Bayesian inference method for mixed Gaussian phylogenetic models, a class of models for PCMs from the $\mathcal{G}_{LInv}$ family of evolutionary processes. This offers flexibility to have models with distinct processes with their own parameters on different parts of a phylogenetic tree. This method also provides the possibility of providing more biologically-informed priors on the parameters.

We demonstrated how prior information can be incorporated into an MGPM, and how the method can appropriately infer the posterior distribution. In addition, we showed that the posterior predictive loss score can be utilized to evaluate the goodness-of-fit of a model and compare several competing hypotheses.

Implementation to the real-world data shows that our method can be used to compare different scenarios of evolution that resulted in the observed data from different taxa that share an evolutionary history.

The natural extension of our method would include multivariate traits that can interact and affect each other throughout their evolution. This extends to the need for a more efficient sampling algorithm to sample from high-dimensional posterior distributions that can result from high-dimensional traits that could possibly include molecular data.

### Acknowledgements
The authors thank Niklas Wahlberg, Masahito Tsuboi, and Sridhar Halali for the valuable discussions, and two anonymous reviewers for their constructive comments.

### About this supplement
This article has been published as part of 19th Conference on Computational Intelligence methods for Bioinformatics and Biostatistics (CIBB), Volume 27 Supplement S1, 2026: 19th Conference on Computational Intelligence Methods for Bioinformatics and Biostatistics. The full contents of the supplement are available at https://link.springer.com/journal/12859/volumes-and-issues/27-1/supplement.

### Author contributions
BB formulated the research project under the guidance of KB. BB developed the method and the R package. BB wrote the first draft of the manuscript. All authors contributed to the writing and proofreading of the manuscript. EY helped with the R package testing and provided insight on the biological side.

### Funding
The authors are supported by the ELLIIT Call C Grant: Developing core-technologies for tree-based models.

### Data availability
The method is available as an R package `bgphy`, available at https://github.com/bayubeta/bgphy. The supporting files for the simulation study and real-world application are available at https://github.com/bayubeta/bgphy-support. The phylogenetic tree of Anolis lizards used for simulation study is available in the supplementary materials of [31]. We scaled the Anolis lizards tree to unit height and made it available in the `bgphy` package as the object `lizardTree`. The antlers dataset and phylogeny for the real-world application are available upon request from the corresponding author of [5].

## Declarations

### Ethics approval and consent to participate
Not applicable.

### Consent for publication
Not applicable.

### Competing interest
The authors declare that they have no Conflict of interest.